\documentclass[12pt,preprint]{aastex}
\usepackage{epsf}

\shortauthors{}
\shorttitle{}

\received{}
\accepted{}
\begin{document}

\newcommand{\mobs}{M_{obs}}
\newcommand{\mtrue}{M_{true}}
\newcommand{\mtobs}{{\tilde m_{obs}}}
\newcommand{\mhat}{{\rm {\hat M}}}
\newcommand{\mttrue}{{\tilde m_{true}}}
\newcommand{\hbeta}{H$\beta$ }
\newcommand{\mg}{\ion{Mg}{2} }
\newcommand{\ca}{\ion{C}{4} }
\newcommand{\mbh}{${\rm M_{BH}}$ }
\newcommand{\etal}{et al.}
\newcommand{\bh}{{\rm BH}}
\newcommand{\bol}{{\rm bol}}
\newcommand{\edd}{{\rm Edd}}
\newcommand{\kms}{{\rm km\;s^{-1}}}
\newcommand{\mmbh}{M_{\rm BH}}

\title{Black Hole Masses and Eddington Ratios at $0.3<z<4$\footnote{Observations reported here were obtained at the MMT Observatory (MMTO), a joint facility of the University of Arizona and the Smithsonian Institution.}}

\author{Juna A. Kollmeier\altaffilmark{2}, Christopher A. Onken\altaffilmark{2,3}, Christopher S. Kochanek\altaffilmark{2}, Andrew Gould\altaffilmark{2}, David H. Weinberg\altaffilmark{2}, Matthias Dietrich\altaffilmark{2},
Richard Cool\altaffilmark{4}, Arjun Dey\altaffilmark{5}, Daniel
J. Eisenstein\altaffilmark{4}, Buell
T. Jannuzi\altaffilmark{5}, Emeric Le Floc'h\altaffilmark{4}, Daniel
Stern\altaffilmark{6}}\altaffiltext{2}
{Dept. of Astronomy, The Ohio State University, 
140 W. 18th Ave, Columbus, OH 43210}
\altaffiltext{3}
{Herzberg Institute of Astrophysics, 5071 West Saanich Road, Victoria, BC V9E~2E7, Canada}
\altaffiltext{4}
{Steward Observatory, University of Arizona, 933 N. Cherry Avenue
Tucson, AZ 85721}
\altaffiltext{5}
{National Optical Astronomy Observatory, P.O. Box 26732, Tucson, AZ, 85719}
\altaffiltext{6}
{Jet Propulsion Laboratory, California Institute of Technology, MS 169-506, Pasadena, CA, 91109 }
\email{jak, onken, ckochanek, gould, dhw, dietrich@astronomy.ohio-state.edu, rcool,elefloch@as.arizona.edu, dey,jannuzi@noao.edu, eisenste@cmb.as.arizona.edu, stern@thisvi.jpl.nasa.gov}

\begin{abstract}
  We study the distribution of Eddington luminosity ratios,
  $L_\bol/L_\edd$, of active galactic nuclei (AGNs) discovered in the
  AGN and Galaxy Evolution Survey (AGES).  We combine H$\beta$,
  \ion{Mg}{2}, and \ion{C}{4}\ line widths with continuum luminosities
  to estimate black hole (BH) masses in 407 AGNs, covering the
  redshift range $z\sim 0.3-4$ and the bolometric luminosity range
  $L_\bol \sim 10^{45}-10^{47}$~erg~s$^{-1}$.  The sample consists of
  X-ray or mid-infrared (24$\micron$) point sources with optical
  magnitude $R\leq 21.5$ mag and optical emission line spectra
  characteristic of AGNs.  For the range of luminosity and redshift
  probed by AGES, the distribution of estimated Eddington ratios is
  well described as log-normal with a peak at $L_\bol/L_\edd \simeq$
  1/4 and a dispersion of 0.3~dex.  Since additional sources of
  scatter are minimal, this dispersion must account for contributions
  from the scatter between estimated and true BH mass and the scatter
  between estimated and true bolometric luminosity.  Therefore, we
  conclude that: (1) neither of these sources of error can contribute
  more than $\sim$0.3~dex rms; and (2) the true Eddington ratios of
  optically luminous AGNs are even more sharply peaked.  Because the
  mass estimation errors must be smaller than $\sim$0.3~dex, we can
  also investigate the distribution of Eddington ratios at fixed BH
  mass. We show for the first time that the distribution of Eddington
  ratios at fixed BH mass is peaked, and that the dearth of AGNs at a
  factor $\sim 10$ below Eddington is real and not an artifact of
  sample selection.  These results provide strong evidence that
  supermassive BHs gain most of their mass while radiating close to
  the Eddington limit, and they suggest that the fueling rates in
  luminous AGNs are ultimately determined by BH self-regulation of the
  accretion flow rather than galactic scale dynamical disturbances.
\end{abstract}


\clearpage

\section{Introduction}

For well over 30 years, the basic theory of active galactic nuclei (AGNs)
has been that they are luminous because of the accretion of matter onto
black holes (BHs; \citealt{salpeter64,zeldovich64,lyndenbell69}).  In this
picture, the luminosity produced by a BH of mass $M_\bh$ has a natural
maximum, the Eddington limit ($L_{\edd}$), at which the radiation pressure
due to the accretion of the infalling matter balances the gravitational
attraction of the BH.  Most models for AGNs assume that they are BHs
radiating near the Eddington limit, and as techniques for independently
estimating BH masses in AGNs have been developed, it has become possible to
test this supposition.  In particular, large, modern spectroscopic surveys
can provide estimates of the Eddington ratios (the ratio of the AGN
bolometric luminosity to the Eddington limit) for thousands of AGNs (see,
e.g., analyses of the Sloan Digital Sky Survey [SDSS] by McLure \& Dunlop
[2004] and M.\ Vestergaard et al., in preparation).  Unfortunately, the
shallowness of these large, wide-area surveys imposes severe restrictions on
the combinations of Eddington ratio and BH mass that are observable,
especially at $z>1$.  For $\mmbh \la 10^9 M_\odot$, the SDSS is sensitive
only to near-Eddington radiators above this redshift, and even at $z<1$ the
SDSS analyses to date have not clearly established whether there is a lower
cutoff to the $L_\bol/L_\edd$ distribution at fixed BH mass.
\citet{warner04} derived BH masses for over 500 ($0 \leq z \leq 5$) AGNs and
found a broad range of Eddington ratios.  However, it is difficult to draw
any conclusions about the underlying distribution of $L_{\bol}/L_{\edd}$
because their dataset is heavily weighted toward high-luminosity objects.

The AGN and Galaxy Evolution Survey \cite[AGES;][] {kochanek04} probes
nearly a decade further down the AGN luminosity function than the SDSS.  For
the first time, this permits a relatively unbiased measurement of the
distribution of Eddington ratios at fixed BH mass at $z\geq 1$, while the
distribution at fixed luminosity can be measured down to $z=0.5$.  The AGES
survey uses the 300-fiber Hectospec robotic spectrograph on the MMT
\citep{fabricant98, roll98, fabricant05} to survey galaxies and AGNs in the
Bo\"otes field of the NOAO Deep Wide-Field
Survey\footnote{\url{http://www.noao.edu/noao/noaodeep/}}
\cite[NDWFS;][]{jannuzi99}.  Both a population of high mass BHs radiating
significantly below Eddington and a population of low mass BHs radiating
near or above Eddington would be observable with AGES.  In this paper we
show: (1) that the distribution of Eddington ratios at fixed BH mass or at
fixed luminosity is narrowly peaked and well-described by a single
log-normal distribution independent of redshift and luminosity, (2) that
this peak occurs at roughly 1/4 of the Eddington limit, and (3) that the rms
error in BH mass estimates from emission-line scaling relations is less than
$\sim$0.3~dex at fixed luminosity.  The first two conclusions imply that the
luminous growth of BHs over cosmic time is dominated by objects radiating
near the Eddington limit.

We present a brief overview of our data from the AGES survey in
\S~\ref{sec:data} and describe our method of analysis in
\S~\ref{sec:analysis}.  We present our BH mass estimates and Eddington
ratios as functions of redshift and luminosity in \S~\ref{sec:results}.
Finally, we discuss the implications of these results in
\S~\ref{sec:discuss}.  In our analysis, we assume an $H_0=72\; \kms\;{\rm
  Mpc^{-1}}, \Omega_m=0.3, \Omega_{\Lambda}=0.7$, flat cosmology.

\section{Data \label{sec:data}}

AGES is a redshift survey in the roughly 9 deg$^2$ Bo\"otes Field that had
been imaged by the NDWFS in $B_{\rm W}$, $R$, and $I$ filters. Subsequent
surveys have imaged the field at many wavelengths, providing a rich
multi-wavelength data set.  In this paper, we make particular use of data
from the {\it Chandra X-ray Observatory} XBo\"otes survey
\citep{murray05,brand06,kenter05}, and the {\it Spitzer Space Telescope}
(MIPS\footnote{The {\it Spitzer} MIPS survey of the Bootes region was
  obtained using Guaranteed Time Observations provided by the {\it Spitzer}
  Infrared Spectrograph Team (James Houck, P.I.)  and by M. Rieke.}: Rieke
et al. 2004; IRAC: Eisenhardt et al. 2004).

The AGES-I Survey (C.\ Kochanek et al., in preparation) selected AGN
candidates as either X-ray or 24$\micron$ sources, with $R \leq 21.5$ mag
(Vega) optical point source counterparts. Objects were considered to be
point sources, if they were point-like in any one of $R$, $I$, or $B_{\rm
  W}$.  Since the luminosities we consider here are higher than the
canonical Seyfert luminosities, there should be little contribution from
host galaxies in these AGNs.  The primary sample of $z>1$ AGNs consists of
either XBo\"otes sources with $\geq 4$ counts (over an average exposure time
of 5 ks) or 24$\micron$ sources brighter than 1~mJy that are off the stellar
locus (2MASS~$J>[12 + 2.5 \;{\rm log} F_{24\mu {\rm m}}]$).  These are
supplemented by X-ray sources with 2 or 3 counts and 24$\micron$ sources
with $0.5~{\rm mJy}<F_{24\mu {\rm m}}<1~{\rm mJy}$.  Redshifts were obtained
with Hectospec at the MMTO.  The spectra were analyzed by two independent
pipelines and verified by eye.  This led to a sample of 733 broad-line AGNs
with $z>0.1$.  We analyze a subset of this sample for which we can reliably
measure emission line widths, as described in \S\ref{sec:fwhm}.

\subsection{Completeness \label{sec:completeness}}

There are three issues for understanding the completeness of our sample: the
literal completeness of the spectroscopy, the effects of the optical flux
limits, and the effects of the 24$\micron$/X-ray flux limits.  The first,
the completeness of the spectroscopy, plays no role in our conclusions.
While we did not obtain spectra of every candidate matching our selection
criteria, we did obtain redshifts for 97\% of the candidates for which we
obtained spectra.  These represented only 66\% of the candidates, but the
candidates with spectra can be regarded as a ``random'' sub-sample of the
candidates that was dictated by the fraction of the NDWFS region covered by
spectroscopy and whether it was possible to assign fibers to the candidates.
Presumably neither of these issues are correlated with either Eddington
ratio or BH mass.  Thus, the only question is whether the spectrum allowed
the measurement of the line FWHM, and we discuss this in \S~\ref{sec:fwhm}.
The second issue, the optical flux limit imposed for the spectroscopic
targets, we will include explicitly in our analysis.

\begin{figure*}
\centerline{ 
\epsfxsize=7.0truein 
\epsfbox[21 148 583 712]{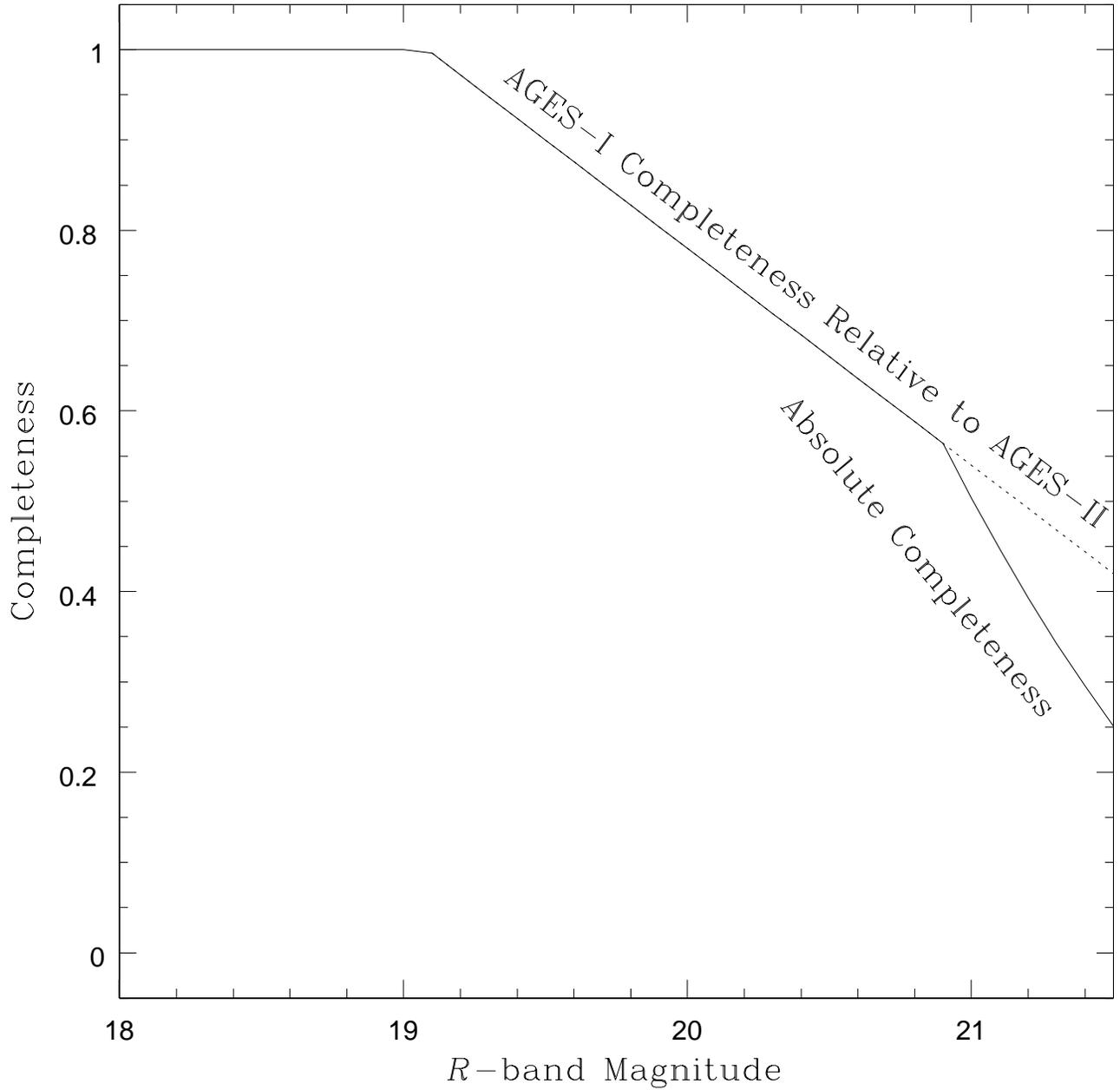} 
}
\caption{Completeness correction as a function of $R$-band magnitude. The
  solid line shows our overall completeness correction. The dashed line
  shows the AGES-I completeness relative to AGES-II.  The departure between
  the two lines shows where we make an additional correction for loss of
  blue objects due to the AGES-II 3.6$\micron$ flux limit. }
\label{fig:completeness}
\end{figure*}

It is the third issue, the consequences of the 24$\micron$ and X-ray
flux limits that we must consider in more detail. Our X-ray limit is
the deepest in the survey \cite[see][]{brand06}, and therefore no
correction is necessary for the X-ray completeness. However, if we
examine the distribution of AGNs in the plane of the $R$-band and
24$\micron$ fluxes, it is clear that both of these flux limits matter.
Our spectroscopic limit is nominally defined by $R = 21.5$ mag.
However, the 24$\micron$+X-ray selected sample is only complete to $R
= 19.1$ mag.  We evaluate the problem and estimate a correction by
using the deeper samples from the AGES-II catalogs (C. Kochanek et
al., in preparation).  AGES-II includes AGNs selected using
mid-infrared colors from the IRAC Shallow Survey \citep{eisenhardt04},
based on the color selection method outlined in \cite{stern05}, as
well as the 24$\micron$ and X-ray criteria used for AGES-I.  This
leads to an AGN sample with a 3.6$\micron$ (approximately $L$-band)
flux limit of $[3.6] = 18$ mag and an optical limit of $I= 21.5$ mag
that fills in most of the missing quasars to the optical flux limit of
AGES-I.  The AGES-II $[3.6] = 18$ mag limit, combined with the fact
that $z>1$ AGNs have $R-[3.6] > 2.9$, implies that the survey is
complete to $R=20.9$ mag.  Brighter than this magnitude, we simply
compare the AGES-I and AGES-II samples to determine the required
completeness correction.  For fainter AGNs, we must apply an
additional correction that incorporates the fraction of AGNs that are
lost in AGES-II owing to their relatively blue colors with respect to
the [3.6] limit.  This correction factor is known from the brighter
magnitudes where AGES-II is complete.  We show the functional form for
our completeness correction in Figure~\ref{fig:completeness}.

These corrections for completeness are important for the interpretation of
the results in \S\ref{sec:mass}, and we return to the completeness
corrections and how they affect our conclusions there.

\section{Analysis \label{sec:analysis}}

BH masses have been estimated from H$\beta$\ emission line widths and
luminosities for some time \cite[e.g.,][]{dibai80}. However, it was
only with the application of reverberation mapping \cite[for an
overview, see][]{peterson01} that this relationship became firmly
established, and then was revised with improved data and techniques
\citep{kaspi96,wandel99,kaspi00,mcj02,vest02,kaspi05,vest06,bentz06}. The
general form of the relation is:
\begin{equation}
\log M_{\rm BH} = a + b \log(\lambda L_{44}) + 2 \log V,
\label{eqn:mhat}
\end{equation}
\noindent where $M_{\rm BH}$ is the estimated BH mass in units of
M$_{\odot}$, $V$ is the H$\beta$\ full width at half maximum (FWHM) in
$\kms$, and $\lambda L_{44}$ is the continuum luminosity near the line (5100
\AA) in units of 10$^{44}\;{\rm erg\;s^{-1}}$. Because optical spectra can
only probe H$\beta$\ to a maximum redshift of $z\sim 0.75$, studies have
been undertaken to find scaling relationships for UV lines, allowing the
estimation of black hole masses at high redshift.  A relationship for
\ion{C}{4}~$\lambda$1549 and the 1350~\AA\ continuum (probing $1.6 \leq z
\leq 5$) was calibrated from reverberation-based H$\beta$\ measurements and
single-epoch \ion{C}{4}\ observations \citep{vest02}, which is well-matched
to the luminosity range probed by AGES-I.  For intermediate redshifts ($0.4<
z < 2$), \citet{mcj02} determined a scaling relationship for
\ion{Mg}{2}~$\lambda$2800 by comparing single-epoch measurements of
\ion{Mg}{2}\ FWHM and 3000~\AA\ continuum luminosity with results from
H$\beta$\ reverberation mapping under the assumption that the two lines are
emitted at the same distance from the BH due to their similar ionization
potentials. While the response of \ion{Mg}{2}\ in reverberation mapping
campaigns has been rather weak (see Clavel et al. 1991 and Dietrich \&
Kollatschny 1995 for the best result, that of NGC~5548), reverberation
studies of \ion{C}{4}\ have yielded BH masses in good agreement with those
from H$\beta$ \cite[e.g.,][]{op02}.  There has been a long struggle to
measure BH masses from these emission lines and each line has associated
peculiarities.  We refer the reader to the original works, which address
these issues in greater detail.

We adopt ($a,~b$)=(0.68,~0.61) in equation~(\ref{eqn:mhat}) for the H$\beta$
relation \cite[from][although other versions of the relation differ by less
than $\approx$0.10 in slope]{mcj02}; we use ($a,~b$)$=$(0.20,~0.7) for
\ion{C}{4}\ \cite[from][]{vest02}; and we estimate in \S~\ref{sec:mg2} that
($a,~b$)$=$(0.31,~0.88) for \ion{Mg}{2}.  We measure the line widths from
the AGES-I spectra (\S~\ref{sec:fwhm}) and the continuum luminosity from the
NDWFS photometry (\S~\ref{sec:contlum}).  We discuss the sensitivity of our
results to the exact parameters of these relations in \S~\ref{sec:mg2fwhm}.

\subsection{Line Width \label{sec:fwhm}}

\begin{figure*}
\centerline{
\epsfxsize=6.5truein
\epsfbox[18 146 581 703]{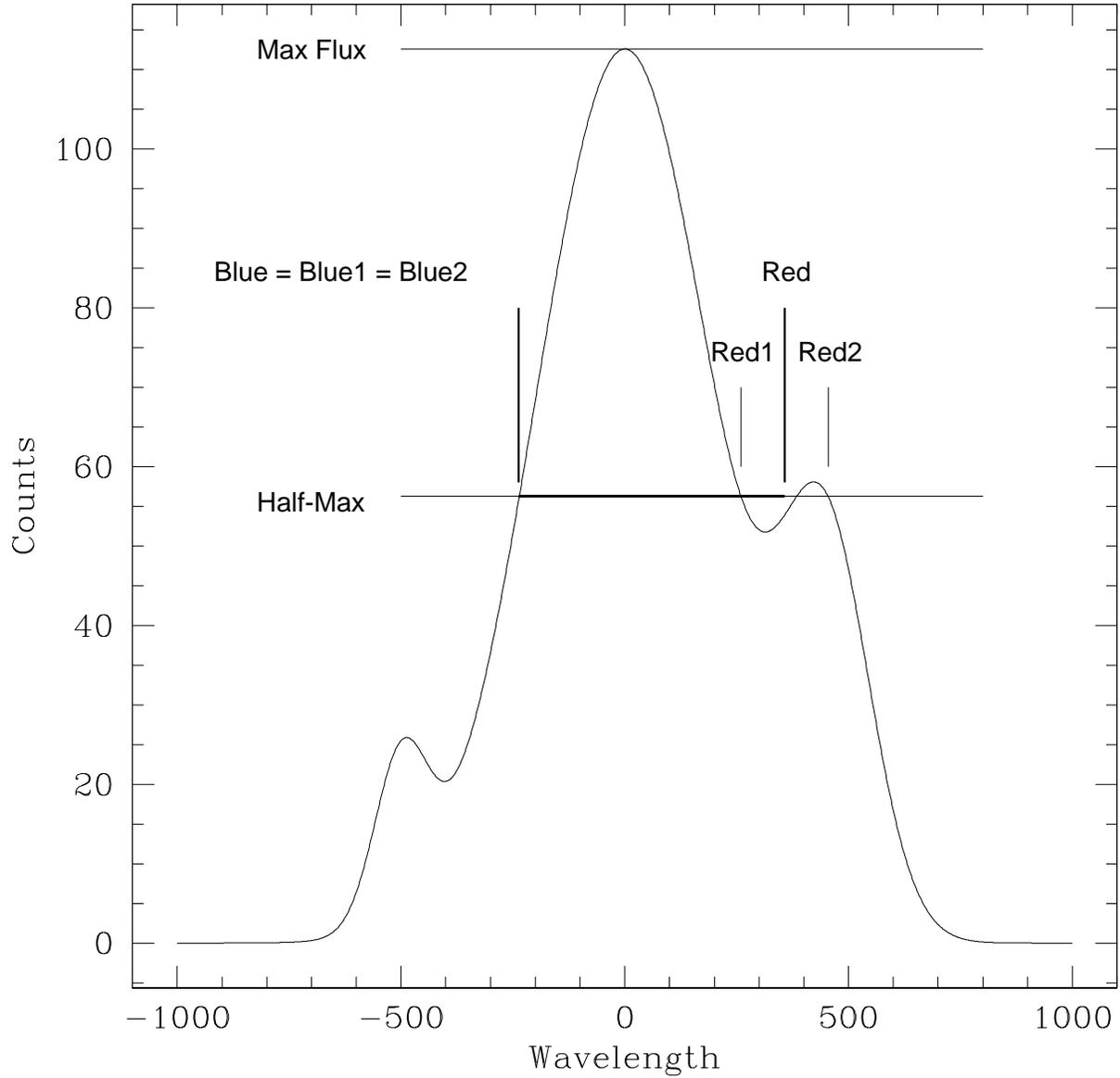}
}
\caption{Illustration of the method for measuring the FWHMs. Blue1 and Red1
  indicate the wavelengths of the first crossing of the half-max point
  descending from the line peak, and Blue2 and Red2 show the first half-max
  crossing ascending from the line limits. Red1 and Red2 are averaged to
  produce the wavelength Red. In this cartoon, Blue1 and Blue2 are identical
  and so are equal to Blue.  The FWHM is determined by the difference
  between Red and Blue.}
\label{fig:demo}
\end{figure*}

The AGES-I spectra have a pixel scale of 1.2~\AA\ and a resolution of
$\approx 6$~\AA\ FWHM.  For our analysis, we boxcar-smooth the spectra over
11 pixels, then subtract a locally-defined linear continuum from the region
around the emission line of interest.  The rest-frame wavelength regions
used to set the continuum around each line are (4740-4765~\AA,
5075-5100~\AA) for H$\beta$, (2670-2682~\AA, 2940-2970~\AA) for \ion{Mg}{2},
and (1455-1465~\AA, 1700-1705~\AA) for \ion{C}{4}.  We subtract narrow-line
contributions to H$\beta$\ using the [\ion{O}{3}]~$\lambda$5007 line as a
model, with the [\ion{O}{3}] flux scaled by a factor of 0.15. This fiducial
value for the (narrow H$\beta$)-to-[\ion{O}{3}] flux ratio lies within the
range defined by local AGNs that have only narrow emission lines (Baldwin et
al.\ 1981; Veilleux \& Osterbrock 1987).  Visual inspection of the
post-subtraction spectra revealed little residual narrow-line contribution.
We determine the peak flux in the line region and measure the FWHM of the
line as follows \cite[adapted from][and illustrated in
Fig.~\ref{fig:demo}]{peterson04}. First, we determine two wavelengths on
each side of the line: (1) the first crossing of the half-max flux level
moving downward from the line peak (indicated as Blue1 and Red1 in
Fig.~\ref{fig:demo}); (2) the first half-max crossing moving upward from the
line edge (Blue2 and Red2). The mean of these two wavelengths is taken as
the half-max point for that side of the line (Blue and Red).  The FWHM is
defined as the difference between the Blue and Red points. The boxcar
smoothing of 13~\AA\ serves to minimize the sensitivity of our automated
procedure to noise in the spectra and results in a final resolution of $\sim
500~\kms$.  This adds a negligible ($\sim$2\%) systematic contribution to
our measured FWHM, for which we do not correct.  We estimate the error in
the FWHM with the technique described by Corbett et al.~(2003), in which
line gradients at the half-max points are used to convert the Poisson noise
of the spectral counts at half-max into wavelength uncertainties. A minimum
FWHM error of 10\% is imposed on all measurements.

To remove spurious measurements from the dataset, every spectrum is examined
for a series of problems: anomalous features (e.g., sky subtraction
artifacts), significant absorption in the line profile, strong \ion{Fe}{2}\
emission around \ion{Mg}{2}, and low signal-to-noise ratio (S/N). The
presence of broad absorption in AGNs has been found to be independent of
Eddington ratio (Scoggins et al. 2004), so the lines removed for strong
absorption (mainly \ion{C}{4}) should be unbiased with respect to Eddington
ratio.  In addition, the Eddington ratio distributions we observe for
objects with masses from \ion{Mg}{2} and from \ion{C}{4} are similar,
despite \ion{Fe}{2} contamination being much less significant near
\ion{C}{4}.  We take this smooth transition as evidence that our cautious
removal procedure only leaves objects for which the FWHM measurements are
secure.

\begin{figure*}
\centerline{ 
\epsfxsize=7.0truein 
\epsfbox[21 148 583 710]{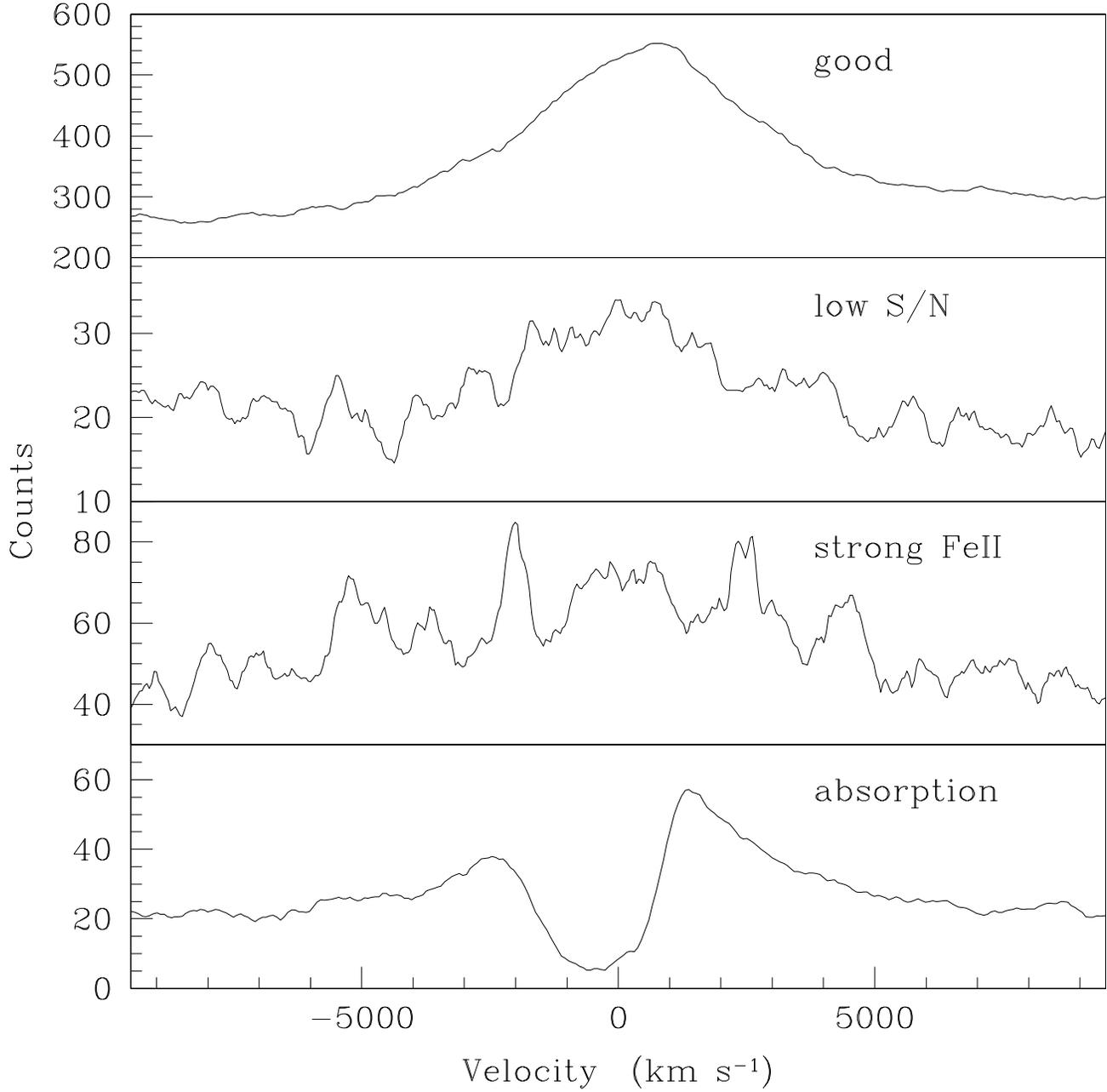} 
}
\caption{Examples of typical smoothed spectra.  Panels from top to bottom
  show examples of an acceptable spectrum, a spectrum rejected for low S/N,
  a spectrum with strong \ion{Fe}{2} emission, and a spectrum with strong
  absorption. }
\label{fig:4spec}
\end{figure*}

We remove 326 of the 733 AGNs in this process: 70 spectral anomalies, 130
for absorption features, 77 with problematic \ion{Fe}{2}\ emission, and 49
with low spectroscopic S/N.  Representative spectra for the final three
classes (and one good spectrum) are shown in Figure~\ref{fig:4spec}. Some
spectra fall into multiple categories, and for our accounting they are
assigned to the first class in this ordering of problems.

Our final data set is composed of 407 AGNs, with mass measurements for (26,
267, 114) AGNs determined from (H$\beta$, \ion{Mg}{2}, \ion{C}{4}).  The
  full range of measured line widths is $1200-5300~\kms$\ for H$\beta$,
  $1400-9700~\kms$\ for \ion{Mg}{2}, and $1800-10600~\kms$\ for \ion{C}{4}.

\subsection{Continuum Luminosity \label{sec:contlum}}

Our Hectospec observations were made during the instrument's inaugural
season of operation.  During our runs, the atmospheric dispersion corrector
did not function consistently, making it difficult to accurately flux
calibrate our spectra. We therefore estimate the continuum luminosities
required for the $M_\bh$ estimates from the broad-band magnitudes.  Using
the 6$^{\prime \prime}$-aperture NDWFS-DR3 $R$ magnitude, we calculate the
flux at the band's effective wavelength (6515~\AA), then compute the
rest-frame 5100, 3000, or 1350~\AA\ fluxes assuming a power-law continuum
with F$_{\lambda} \sim \lambda^{-1.7}$.  This spectral index is the average
spectral slope measured for the $\sim$700 AGNs of comparable luminosity in
the sample of \citet{dietrich02}.  The range around the median slope is less
than 0.15 (M.~Dietrich \& F.~Hamann, in preparation), so our adoption of a
fixed slope contributes little, on average, to our total observational
uncertainties. We note that the Dietrich \& Hamann measurements are
consistent with the results from the SDSS composite spectra
\citep{vandenberk01}.  Contributions of emission lines to the $R$-band flux
should be less than 0.2 mag.

\subsection{Calibrating the \ion{Mg}{2} Relation \label{sec:mg2}}

Initially, we used the \ion{Mg}{2} relation of \citet{mcj02}, for which
($a,~b$)=(0.53,~0.47) in equation~(\ref{eqn:mhat}).  However, for AGNs in
the redshift ranges in which we could estimate $M_\bh$ using both
\ion{Mg}{2} and a second line ($0.4 \leq z \leq 0.75$ for H$\beta$ and $1.6
\leq z \leq 2.0$ for \ion{C}{4}), we found systematic differences when using
this relation.  As shown in Figure~\ref{fig:dm.l}, there is a linear trend
in [$\log M_\bh$(\ion{Mg}{2})$ - \log M_\bh$(H$\beta$ or \ion{C}{4})] with
$\log L_\bol$ that has a similar slope for both redshift ranges/lines.
Fitting for the luminosity trend, we find a consistent relationship for both
H$\beta$ and \ion{C}{4}, ensuring that all three lines will be on a common
scale.  We redefine the \ion{Mg}{2} relation from this fit and find that
($a,~b$)=(0.31,~0.88), while the \citet{mcj02} slope is ruled out at the
5-$\sigma$ level.  Although we use this modified relation in the current
work, we are not at this time advocating our revised relation for other
studies because of the differences between our analysis and that presented
in the \citet{mcj02} paper.  After adjusting the slopes, the residual
scatter between \ion{Mg}{2} and H$\beta$ or \ion{C}{4} mass estimates
indicates an intrinsic dispersion beyond the measurement errors of 0.27~dex.
This is less than the 0.4~dex of intrinsic scatter found in earlier studies
of \ion{Mg}{2} (McLure \& Dunlop 2004) and suggests that \ion{Mg}{2} $M_\bh$
estimates are of comparable accuracy to those from H$\beta$ or \ion{C}{4},
although in principle all three mass relations could be affected by the same
zero-point error.

In addition to the scaling with luminosity, we examined the mass difference
as a function of \ion{Mg}{2}\ FWHM, allowing both the luminosity and FWHM
scaling to be free parameters in our fit.  We find that the best-fit
relation yields a FWHM scaling of 2.13$\pm$0.34. We interpret the
consistency with the standard virial assumption as a positive indication for
the validity of the \ion{Mg}{2}-based technique.

\subsection{Bolometric Luminosity Calculation}

Following \citet{kaspi00}, we estimate the bolometric luminosity as
$L_{\bol} \simeq 9 \times \lambda L_{\lambda} (5100~{\rm \AA})$, a value
that assumes an AGN spectral energy distribution (SED) typical of optically
selected quasars with little dust obscuration.  We calculate the rest-frame
flux at 5100~\AA\ employing the same method that we used to estimate
continuum luminosities in \S~\ref{sec:contlum}.  While this extrapolation to
5100~\AA\ is quite far for the highest redshift AGNs in our sample,
published conversions to $L_\bol$ using alternative continuum regions affect
the resulting luminosities by less than 30\% (for our assumed spectral
slope).  Of course, departures from the standard AGN SED could change the
relation between $L_\bol$ and $\lambda L_\lambda(5100~{\rm \AA})$, so it is
important to keep in mind that the ``bolometric'' luminosities referred to
throughout this paper are really just multiples of the optical luminosity.

\subsection{Framework}

We emphasize here that we are adopting the theoretical framework that black
hole mass can be reliably measured from the virial relationships. Our
analysis requires knowledge of black hole masses and bolometric
luminosities, but we can only make estimates of these quantities.  Because
there are dispersions in the spectral slopes and bolometric corrections of
real AGNs, our adoption of fixed values for these parameters causes us to
misestimate $L_\bol$.  Intrinsic dispersion in the virial relationships
causes us to misestimate $M_\bh$.  The combination of the intrinsic
dispersion of Eddington ratios and our misestimations of $L_\bol$ and
$M_\bh$ result in the observed scatter of our measurements.

A primary goal of this study is to understand the mass distribution of
active BHs at a fixed luminosity and redshift. The homogeneous nature of our
sample and the depth of the AGES-I survey combine to allow us to address the
underlying distribution over a wider range of $L_\bol$ and $z$ than other
studies.  Figure~\ref{fig:lofz} shows our sample of AGN luminosities as a
function of redshift, along with curves at fixed magnitudes, to illustrate
how the optical flux limits restrict the sample.  We also show how the
AGES-I data are distributed relative to the knee of the luminosity function
as determined by the 2dF-SDSS LRG and QSO Survey
\cite[2SLAQ;][]{richards05}.  In a bright survey like the SDSS, for which
the $i^\prime=19.1$ mag limit (for $z<3$ AGNs) corresponds to $R \approx
18.9$ mag and the $i^\prime=20.2$ mag limit (for $z>3$) corresponds to
$R\approx 20.0$ mag, one cannot compare AGNs of fixed luminosity over as
broad a redshift range as with AGES-I.

\section{Results
\label{sec:results}}

Figure~\ref{fig:seurat} shows the distribution of AGNs in inferred BH mass,
$M_\bh$, and bolometric luminosity, $L_\bol$.  The points are color-coded by
redshift, and we show only a typical error bar to avoid clutter.  These
uncertainties reflect only the {\it statistical} errors in the line width
and luminosity estimates, and we will discuss the possible consequences of
systematic errors in \S~\ref{sec:discuss}.  Figure \ref{fig:seurat} has
several striking features.  First, it is characterized by a fairly narrow
ridge that extends diagonally, with unit slope, across the entire diagram.
Second, this ridge is separated from the solid line representing the
Eddington limit by approximately 0.6~dex.  Third, at a given luminosity, the
density of points falls rapidly toward higher $M_\bh$, i.e., at small
Eddington ratio.  Most of the conclusions of this paper are derived by
quantifying these features.  At fixed $L_\bol$, the ridge is broader than
the statistical uncertainties, indicating that it is dominated either by
intrinsic scatter in the underlying distribution or by scatter between the
inferred values of $M_\bh$ and $L_\bol$ and their true values.  While the
systematic errors in $L_\bol$ should be modest, determining the systematic
errors in $M_\bh$ is a notoriously difficult problem to which we will return
in \S~\ref{sec:discuss}.  However, from the ridge's relatively narrow width
it is already clear that this scatter cannot be too large.  The fact that
the ridge is displaced from the Eddington limit shows that the great
majority of observed AGNs radiate at modestly sub-Eddington rates, assuming
that our $M_\bh$ calibration is correct in the mean.  The fact that there
are few AGNs above the ridge implies that the observed (in practice,
luminosity-limited) AGNs are dominated by BHs radiating close to Eddington
rather than more massive BHs radiating well below Eddington.  The data
presented in Figure \ref{fig:seurat} also permit one to pose an orthogonal
question: what is the distribution of AGN luminosities at fixed $M_\bh$?
However, this question cannot be addressed by mere inspection of this figure
and will require additional analysis.

AGNs rejected from the final sample due to low S/N could induce the illusion
of a narrow ridge if these lay preferentially near the selection boundary.
Figure~\ref{fig:seuratbw} shows the $\log L_{\bol}$ vs. $\log M_{\bh}$
positions of the AGNs eliminated due to low S/N spectra and can be compared
to Figure~\ref{fig:seurat}. We have not included in
Figure~\ref{fig:seuratbw} objects rejected due to absorption, \ion{Fe}{2}
emission, or other anomalous features. These objects (for which we show the
$L_{\bol}$ distribution in the inset of Fig.~\ref{fig:seuratbw}) have
unmeasurable FWHMs, and it therefore is not sensible to assign any mass
value, in contrast to AGNs for which we simply have low confidence in the
measured value.  Objects were rejected without reference to their position
in Figure~\ref{fig:seuratbw}, and there is no strong clustering of their
position in this diagram.  The simplest interpretation of this broad, flat
distribution is that the FWHM measurements are simply bad, and that the
calculated $M_{\bh}$ values bear only a casual relation to the true values.
While the inclusion of these uniformly distributed bad measurements would
slightly increase the dispersion of our measurements, we believe that their
rejection is justified.  In any case, including them does not qualitatively
alter our results, as we show in \S~\ref{sec:mass}. Hence, we adopt the
conclusions based on their exclusion.

In the next two subsections we examine the distributions of Eddington ratios
at fixed $L_\bol$ (\S~\ref{sec:lum}) and fixed $M_\bh$ (\S~\ref{sec:mass})
as functions of redshift.

\subsection{Luminosity-Redshift Bins
\label{sec:lum}}

In order to characterize the true distributions of masses and
luminosities for the BHs in our sample, we perform a very simple
maximum likelihood analysis, fitting the data assuming a model in
which the true distribution of $\log(L_{\bol}/L_{\edd})$ is Gaussian
(i.e., $L_\bol/L_\edd$ is log-normal).  We divide the sample into 2
bins in redshift and 3 bins in luminosity.  The redshift division is
at $z=1.2$, and the luminosity cuts are at $10^{45.5}~{\rm erg\;
s^{-1}}$ and $10^{46.0}~{\rm erg\; s^{-1}}$ in both redshift bins. Due
to the different redshift distributions of sources at high and low
luminosity, it is impossible to make cuts that yield similar numbers
of objects in each bin.  We therefore choose cuts such that we have at least
$\sim$10 objects in each bin.  In each $(L_\bol,z)$ bin we estimate
the mean and dispersion of $\log(L_{\bol}/L_{\edd})$, corrected for
the statistical uncertainties.

The histograms in Figure~\ref{fig:lumbin} show the distributions of
estimated $\log(L_\bol/L_\edd)$ in each $(L_\bol,z)$ bin.  For each panel,
we calculate the unweighted mean and standard deviation of the data points,
and we plot a Gaussian with those parameters as a dashed curve in the panel.
Solid curves show the Gaussians with our maximum likelihood fit parameters,
which account for the statistical errors in the linewidth and luminosity of
each data point.  Overall, these statistical errors are small compared to
the widths of the histograms, so there is little difference between the
dashed and solid curves.  Table~\ref{tab:mgauss} lists the measured mean,
dispersion, skewness, and kurtosis of the distributions in each $(L_\bol,z)$
bin, and the mean and dispersion of the maximum likelihood Gaussian fits.

The first point to note from Figure~\ref{fig:lumbin} is that the
distribution of $\log(L_\bol/L_\edd)$ measurements in each bin is nearly
Gaussian, with a center and width that is approximately independent of
redshift and luminosity.  This result quantifies the impression from
Figure~\ref{fig:seurat} that the observed AGNs in our sample are
predominantly BHs radiating fairly near Eddington ($L_\bol/L_\edd\sim 1/4$),
rather than more massive BHs at $L_\bol/L_\edd\ll 1$.  The skewness and
kurtosis of the distributions are in good agreement with the values $S_k=0$
and $A_k=3$ expected for a Gaussian distribution.

The second point to note is that these Gaussians are rather narrow:
$\sigma\approx0.3$~dex after accounting for the statistical errors in our
linewidth and luminosity measurements.  These widths reflect the width of
the intrinsic distribution of Eddington ratios at fixed luminosity {\it and} the errors introduced by our use of universal values for the bolometric
  correction and the spectral slope. While there may be scatter around these
  values or around the relationships used to derive our BH masses and AGN
  luminosities, that scatter already appears in Figure~\ref{fig:seurat}. For
  every object in which some quantity is underestimated by the use of the
  mean relation, another object will have that parameter overestimated.

  The only source of scatter in the estimates of $M_\bh$ that is {\it
  not} fully reflected in the distribution of Eddington ratios is the
  difference between the observed luminosity and the time-averaged
  continuum luminosity in the neighborhood of the broad line for a
  given AGN, the latter being the quantity most appropriate for the
  $M_\bh$ calculation. This is because both $L_\bol$ and $M_\bh$ have
  a luminosity dependence.

  One may show that, if the rms difference between the mean $\log L$ and the
  one derived from the observations is $\sigma_{\log L}$, then the
  dispersion in mass determinations is larger than the observed dispersion
  in Eddington ratios, $\sigma_{\log E}$, by
\begin{equation}
\sigma_{\log M}^2 = \sigma_{\log E}^2 + (2b -1)\sigma_{\log L}^2
\end{equation}
where $b$ is the power-law of the luminosity-dependence in
equation~(\ref{eqn:mhat}). We identify three sources of $\sigma_{\log
L}$: measurement error $\sigma_{\log L,{\rm meas}}\sim0.01$~dex,
scatter due to AGN variability $\sigma_{\log L,{\rm
var}}\sim0.11$~dex, and extrapolation of the line continuum from the
$R$-band measurement $\sigma_{\log L,{\rm
slope}}\sim0.04$~dex\footnote{The amplitudes of these components are
calculated from 1) the typical $R$-band photometric uncertainty, 2)
the rms variability of AGNs over the light-crossing time for the
largest broad-line regions in our sample \cite[i.e., the maximum time
between a luminosity change and the response of the emission line
width;][]{devries05}, and 3) the effect of the 1$\sigma$ error on the
slope extrapolated to the ends of the observed spectrum.}.  Combining
these in quadrature, and using $\sigma_{\log E}=0.30$~dex, we find
$\sigma_{\log M}=0.32$~dex, which differs negligibly from
$\sigma_{\log E}$. We note that this estimate of $\sigma_{\log M}$
represents the maximum possible error on $M_\bh$ and assumes that the
intrinsic Eddington ratio distribution has no width of its own.

The conclusion that the quadrature sum of measurement errors in $M_\bh$,
scatter in bolometric corrections, scatter in spectral slopes, and
variations in $L_\bol/L_\edd$ is only $\sim$0.3~dex rms is rather
remarkable.  The data give no clear indication of how to partition the
scatter between these contributions. The rms scatter of 0.3~dex in inferred
$M_\bh$ seems plausible on geometrical grounds, since the relation between
observed linewidth and BH mass may depend on viewing angle
(\citealt{krolik01,collin06}; C.\ Kuehn \& B.\ Peterson, in preparation). If the
observational errors or the distribution of viewing angles do dominate the
width of the histograms, then the intrinsic distribution of $L_\bol/L_\edd$
is very narrow indeed.

Figure~\ref{fig:monte_lumbin} compares the observed distribution in the
$\log L_\bol - \log M_\bh$ plane (shown earlier in Fig.~\ref{fig:seurat}) to
three Monte Carlo realizations in which we draw an estimated BH mass for
each observed AGN from a Gaussian in $\log(L_\bol/L_\edd)$ with mean of
$-0.6$~dex and dispersion of 0.3~dex.  It is clear by visual inspection that
this simple model describes the observed distribution of points remarkably
well.  While one may be tempted to see outliers or gaps in
Figure~\ref{fig:seurat}, these features commonly appear in Monte Carlo
realizations of the sample made assuming the log-normal model.

We stress that from an observational standpoint, the procedure followed in
this section is extremely clean.  Since the sample completeness is
well-characterized at fixed optical luminosity and is independent of
Eddington ratio, there are almost no selection effects coming into play in
any of the luminosity bins.  Application of our completeness corrections
have nearly zero effect on these histograms.  On the other hand, {\it
  within} each luminosity bin there is a contribution to the dispersion in
the histogram from uncertainties in estimating $M_\bh$. The robust nature of
binning by luminosity also applies to AGN surveys with other flux limits.
For example, in Figure~3 of \citet{mcd04}, who analyzed more than 12,000
AGNs from SDSS, one can make similar cuts at constant $\dot M_{\rm BH}$
($\propto L_\bol$). \citet{mcd04} calculated the mean Eddington ratio as a
function of redshift, finding the mean to increase slightly with $z$ (from
0.15 at $z\simeq 0.2$ to 0.5 at $z\simeq 2.0$), but they do not address the
shape of the distribution at fixed luminosity, nor, for the redshifts that
they can probe, the distribution at fixed mass.  \cite{woo02} find a much
broader distribution of Eddington ratios in their sample of 234 AGNs
compiled from the literature.  Because their sample consists of low redshift
($z<1$) and local ($z<0.1$) systems, and extends to bolometric luminosities
as low as $10^{43}$~erg~s$^{-1}$, the difference from our results could
plausibly be explained by the very different properties of the samples.  The
\cite{warner04} and \cite{vest04} studies of heterogeneous samples of AGNs
are most akin to what we have presented here.  In both works, Eddington
ratios are presented over a wide range of redshifts and luminosities.
However, because both studies relied on diverse mixtures of samples, the
selection effects are not easily understood, and therefore, the underlying
distribution of Eddington ratios cannot be determined.  For example, the
\citet{warner04} luminosity distribution is heavily weighted towards high
luminosity objects.  Two of the main advantages of the AGES-I survey are the
homogeneity of the sample and the relatively straightforward completeness
corrections (see \S~\ref{sec:completeness}), which allow us to address the
underlying distribution of Eddington ratios.

\subsection{Mass-Redshift Bins
\label{sec:mass}}

The distribution of Eddington ratios at fixed luminosity is a convolution of
the luminosity distribution at fixed BH mass with the underlying BH mass
function (see \citealt{steed03}).  Thus, the fall-off at low Eddington
ratios in Figure~\ref{fig:lumbin} could, in principle, be attributed
primarily to a rapid fall-off in the $M_\bh$ distribution toward higher
masses.  Since we know from our previous discussion that the errors in BH
mass assignments must be fairly small (0.3~dex rms at most), we can bin the
sample by estimated BH mass and look directly at the distribution of
Eddington ratios at fixed $M_\bh$.  Figure~\ref{fig:massbin} shows these
distributions for mass bins $\log(M_\bh/M_\odot)=7-8$, $8-9$, and $9-10$,
and redshift bins $z=1-2$, $2-3$, and $3-4$.

In contrast to Figure~\ref{fig:lumbin}, the distribution of Eddington
ratios at fixed mass is affected by the magnitude limit of the
survey. We also must include the effects of incompleteness at the
faint end. We use the completeness function of
Figure~\ref{fig:completeness} and the $R$-band magnitude of each of
our 407 AGNs to calculate the number of objects at each observed
combination of $L_{\bol}$ and $z$ that are ``missing''. The missing
objects are then distributed in mass at the ($L_{\bol}$,$z$) of the
real AGN via the Eddington ratio distribution we found in
\S~\ref{sec:lum}. For all objects, we take the distribution as
log-normal in $L_\bol/L_\edd$, with a mean of 1/4 and with
$\sigma$=0.30~dex. The contribution of these missing objects to each
bin of Eddington ratio is then summed within the mass and redshift
ranges chosen for each panel.  The solid histograms in each panel are
completeness-corrected.  This alters the histograms, but does not
substantially alter our conclusions.  The vertical arrow in each panel
of Figure~\ref{fig:massbin} indicates the point in that mass-redshift
bin at which AGNs are first lost to optical selection effects (which
begin at the low-$M_\bh$, high-$z$ corner of the bin), and the shaded
region indicates the point at which all AGNs (extending to the highest
$M_\bh$ and lowest $z$ corner of the bin) are lost.  For the
lower-mass bins (particularly at high redshift), the optical magnitude
limit truncates the distributions, and we cannot determine whether
there is a true cutoff at low $L_\bol/L_\edd$.  However, in some bins,
the optical magnitude limit only affects the regime $L_\bol/L_\edd \la
0.1$, and it is possible to test for a real cutoff.  In the
highest-mass highest-redshift bin of Figure~\ref{fig:massbin}, there
are too few objects to make a robust determination.  However,
Figure~\ref{fig:massbingood} shows four bins --- $\log(M_\bh/M_\odot)
= 8.5-10$ at $z=0.75-1$ and $1.5-2$, $\log(M_\bh/M_\odot) = 8.75-10$
at $z=1-1.5$, and $\log(M_\bh/M_\odot) = 9-10$ at $z=2-3$ --- in which
the statistics are relatively good and the fall-off appears to be
real.  The Poisson probabilities that the peak in each case is a
statistical fluctuation are found to be low, as shown in each panel in
Figure~\ref{fig:massbingood}.  The peak is therefore robust and not
merely a product of counting statistics.

Finally, Figure~\ref{fig:massbingood2} shows the effect of including the low
S/N measurements from Figure~\ref{fig:seuratbw} on our Eddington ratio
distributions at fixed mass. The low S/N spectra do not affect any of our
conclusions above.

We note that the scatter in inferred $M_\bh$ might move AGNs from one
mass bin to another, which could potentially enhance the appearance of
a peak either on the rising or the falling sides.  Recall from
\S~\ref{sec:lum} and Table~\ref{tab:mgauss} that the rms scatter
in the inferred $M_\bh$ must be less than $\sim$0.3~dex.  Because the
mass bins in Figure~\ref{fig:massbingood} have widths of at least
1~dex, scatter between mass bins must be a small effect.  The smooth
distribution of Eddington ratios seen in Figure~\ref{fig:seurat}
implies that it would be difficult to import structure into or out of
a particular mass bin without having significant structure in another
part of the distribution from which objects can scatter.  No such
structure is seen in our data. In fact, if we bin the Monte Carlo
realizations of Figure~\ref{fig:monte_lumbin}, which have a
$\sigma=0.3$~dex Gaussian distribution of estimated
$\log(L_{\bol}/L_{\edd})$ and which match the sample's observed
luminosity distribution by construction, then we get histograms that
qualitatively resemble those in the four mass-redshift bins of
Figure~\ref{fig:massbingood}.  This simple model therefore appears
consistent with our data, although this experiment on its own does not
tell us whether the observed distributions are shaped primarily by
intrinsic scatter in $L_\bol/L_\edd$ or by errors in $M_\bh$
estimation.

The filled circles at the bottom of each panel in Figures~\ref{fig:massbin}
and \ref{fig:massbingood} show the relevant spectroscopic limit of the SDSS
for each redshift bin.  The SDSS data are not deep enough to probe the
cutoff in any of these bins, i.e., the SDSS cannot provide an unbiased study
of the Eddington ratio distribution below 0.1 for BHs with $M_\bh < 10^{9}
M_{\odot}$ above $z\sim 1$, or $M_\bh < 10^{9.6} M_{\odot}$ above $z\sim 2$.
Because of its large area, the SDSS can obtain better statistics for high
mass BHs at low redshift, or for very massive BHs at higher redshift, and it
will be interesting to see whether these have a peaked Eddington ratio
distribution similar to that found here.

\subsection{\ion{Mg}{2}\ Scaling \label{sec:mg2fwhm}}

The calibration of \ion{Mg}{2} presents a potential problem for the
conclusions drawn from the AGES-I data that make use of this line.  Our
empirical calibration leads to a steeper dependence of $M_{\bh}$ on
luminosity for the \ion{Mg}{2}\ relationship (\S~\ref{sec:mg2}) than for
\ion{C}{4} or H$\beta$.  As pointed out by \citet{woo02}, if the dependence
of black hole mass on luminosity is nearly linear and there is a simple
relationship between the continuum and bolometric luminosities, then a
strong correlation of increasing $M_\bh$ with increasing $L_\bol$ will be
automatically introduced.  With such a scaling relation, it would be
possible for a small, but random distribution of \ion{Mg}{2}\ FWHMs to
reproduce a tight Eddington ratio distribution without actually being
related to the BH mass.  However, the clear trend in Figure~\ref{fig:dm.l}
and the fact that the \ion{Mg}{2}\ relation scales as $V^{2}$, even when the
velocity dependence is allowed to vary, indicates that the relationship
among the 3000~\AA\ continuum luminosity, the \ion{Mg}{2} line width, and
the black hole mass is {\it not} random.

Our empirical calibration for \ion{Mg}{2}\ depends directly on what we use
for the H$\beta$ and \ion{C}{4} relations.  A new calibration of those two
emission lines has recently been published \citep{vest06}, which has derived
shallower slopes for both the H$\beta$ and \ion{C}{4}\ luminosity scalings,
with ($a, b$) values of (0.91, 0.50) and (0.66, 0.53), respectively.  The
effect of adopting those relations is to shift the H$\beta$ points up
$\sim$0.2~dex in mass (as they have typical luminosities near 10$^{45}$
erg~s$^{-1}$), and to shift the \ion{C}{4}\ points up by $\sim$0.28~dex at
the low luminosity end ($\sim 10^{46}$ erg~s$^{-1}$), and to leave them
nearly unchanged at the high luminosity end.  Using these relations then
sets the parameters of our empirically calibrated \ion{Mg}{2}\ mass relation
to (0.55, 0.82).  This shift produces roughly coherent movement across the
$M_{\bh}-L_{\bol}$ plane, so the shapes of our histograms at fixed mass are
basically unaffected---although our highlighted bins in
Figure~\ref{fig:massbingood} must be shifted upwards by $0.25-0.5$~dex in
mass.

\section{Discussion 
\label{sec:discuss}}

Our survey of $R\leq 21.5$ mag, X-ray- and 24$\micron$-selected AGNs in the
Bo\"otes NDWFS field yields four basic results.  First, the rms scatter in
BH masses inferred from linewidth-luminosity scaling relations is less than
0.3~dex, at least for $M_\bh > 10^8 M_\odot$ and the luminosity range
$L_\bol \sim 10^{45}-10^{47}$~erg~s$^{-1}$.  Second, luminous AGNs at
  $z>0.5$ are powered by BHs radiating at roughly $1/4$ of the Eddington
  rate: there are few cases of higher mass BHs radiating well below
Eddington or of lower mass BHs radiating well above Eddington.  Third, at
{\it fixed mass} above $10^{8.5}~M_\odot$ (where selection effects are
inconsequential), the distribution of Eddington ratios at $z>0.75$ is
strongly peaked.  Fourth, the distribution around this peak is confined to
within $\sim 1$~dex in $L_{\bol}/L_{\edd}$, with many fewer broad-lined AGNs
on either side.

\begin{figure*}
\centerline{
\epsfxsize=6.5truein
\epsfbox[18 146 581 703]{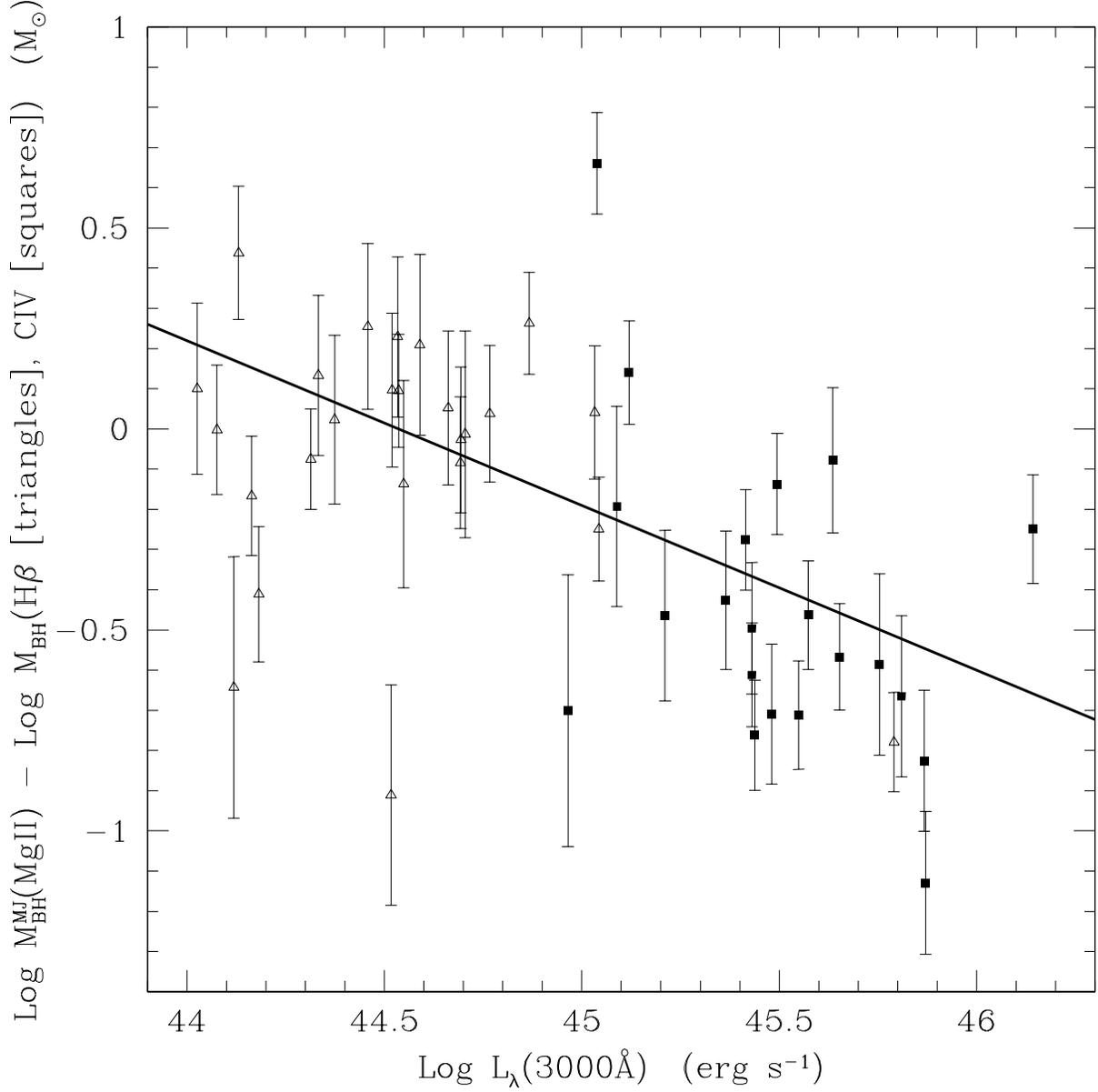}
}
\caption{Comparison of BH masses derived from the \cite{mcj02} \ion{Mg}{2}\
  scaling relation and our adopted H$\beta$ (open triangles) or \ion{C}{4}\
  (filled squares) scaling relations in redshift regimes of overlap, as a
  function of bolometric luminosity, $L_\bol$. The line shows the best fit
  to the combined dataset and forms the basis for our modified calibration
  of the \ion{Mg}{2}\ relation. }
\label{fig:dm.l}
\end{figure*}

\begin{figure*}
\centerline{
\epsfxsize=6.5truein
\epsfbox[18 146 581 703]{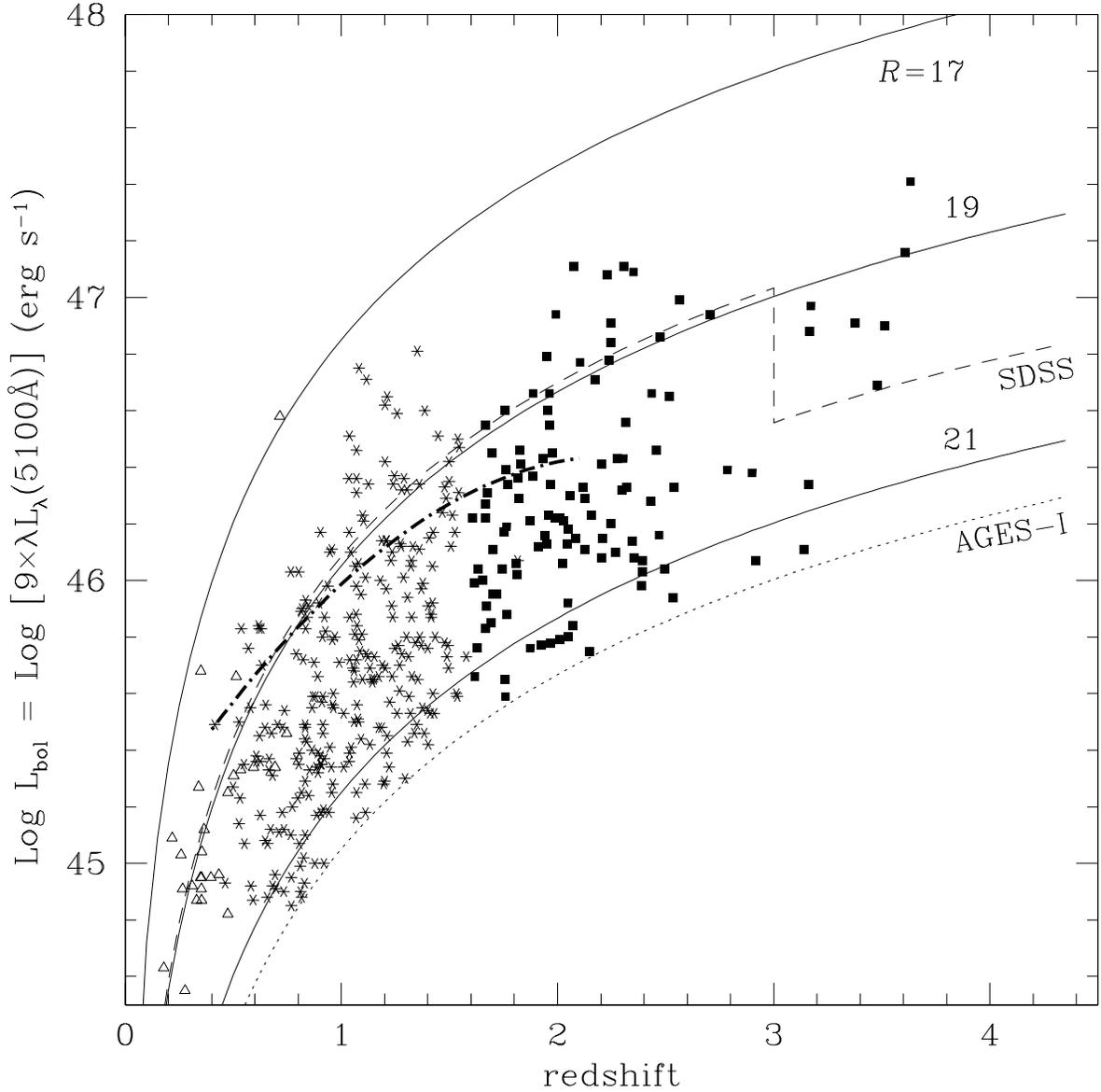}
}
\caption{AGN luminosity as a function of redshift. Points are coded
  according to the emission line used for the mass measurement, with open
  triangles, asterisks, and filled squares corresponding to H$\beta$,
  \ion{Mg}{2}, and \ion{C}{4}, respectively. Also shown are solid curves at
  $R$ = (17, 19, 21)~mag, the SDSS spectroscopic flux limit as a function of
  redshift (dashed line; $z<3$: $R \approx 18.9~{\rm mag}$, $z>3$: $R
  \approx 20.0~{\rm mag}$), and the AGES-I spectroscopic limit (dotted line;
  $R$=21.5~mag).  The dot-dashed line shows the evolution of the knee in the
  luminosity function with redshift, as determined by the 2SLAQ survey
  \citep{richards05}. }
\label{fig:lofz}
\end{figure*}

\begin{figure*}
\centerline{
\epsfxsize=5.5truein
\epsfbox[18 146 581 703]{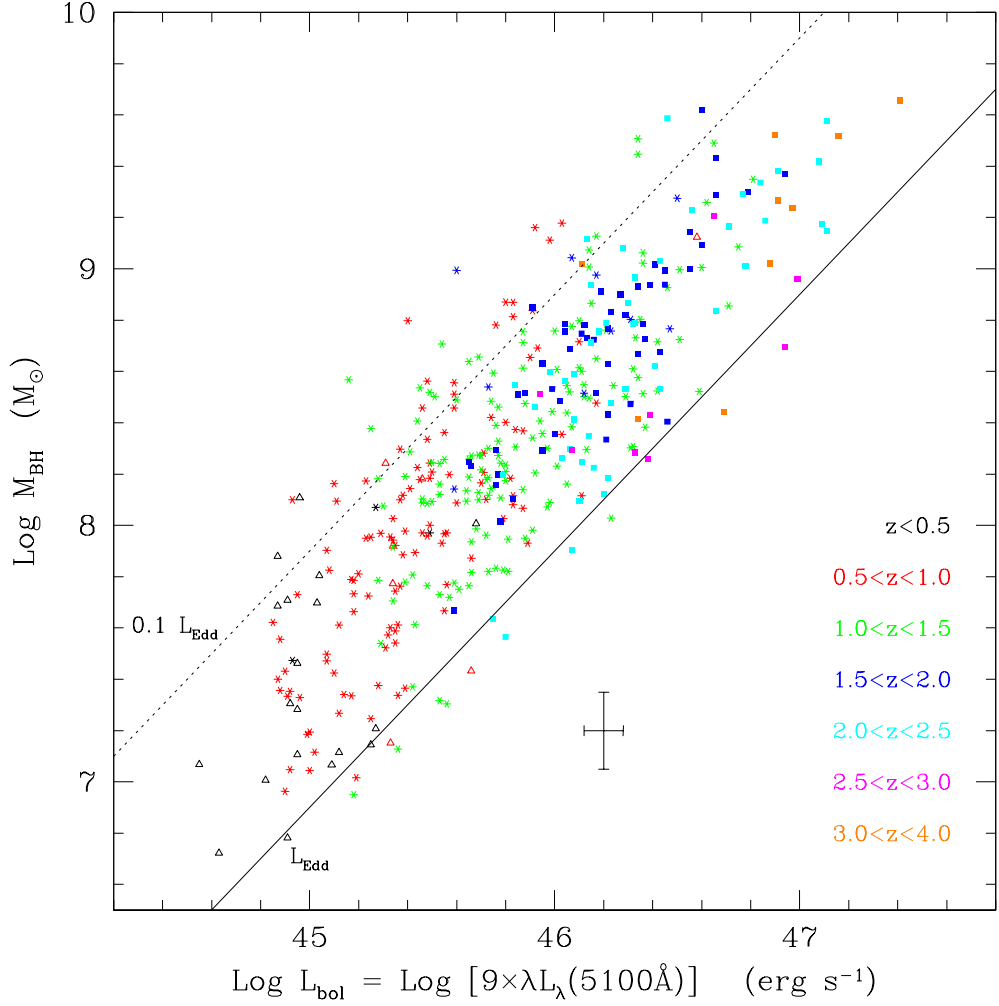}
}
\caption{Estimated BH masses as a function of AGN bolometric luminosity.
  Objects are color-coded by redshift range as indicated in the legend.  The
  solid line denotes the Eddington limit, $L_\edd$; objects to the right of
  the line are radiating above Eddington for the measured mass. The dotted
  line denotes one-tenth of the Eddington limit. Point types denote the
  emission line used for the mass measurement, with open triangles,
  asterisks, and filled squares corresponding to H$\beta$, \ion{Mg}{2}, and
  \ion{C}{4}, respectively.  }
\label{fig:seurat}
\end{figure*}

\begin{figure*}
\centerline{ 
\epsfxsize=6.5truein 
\epsfbox[18 146 581 703]{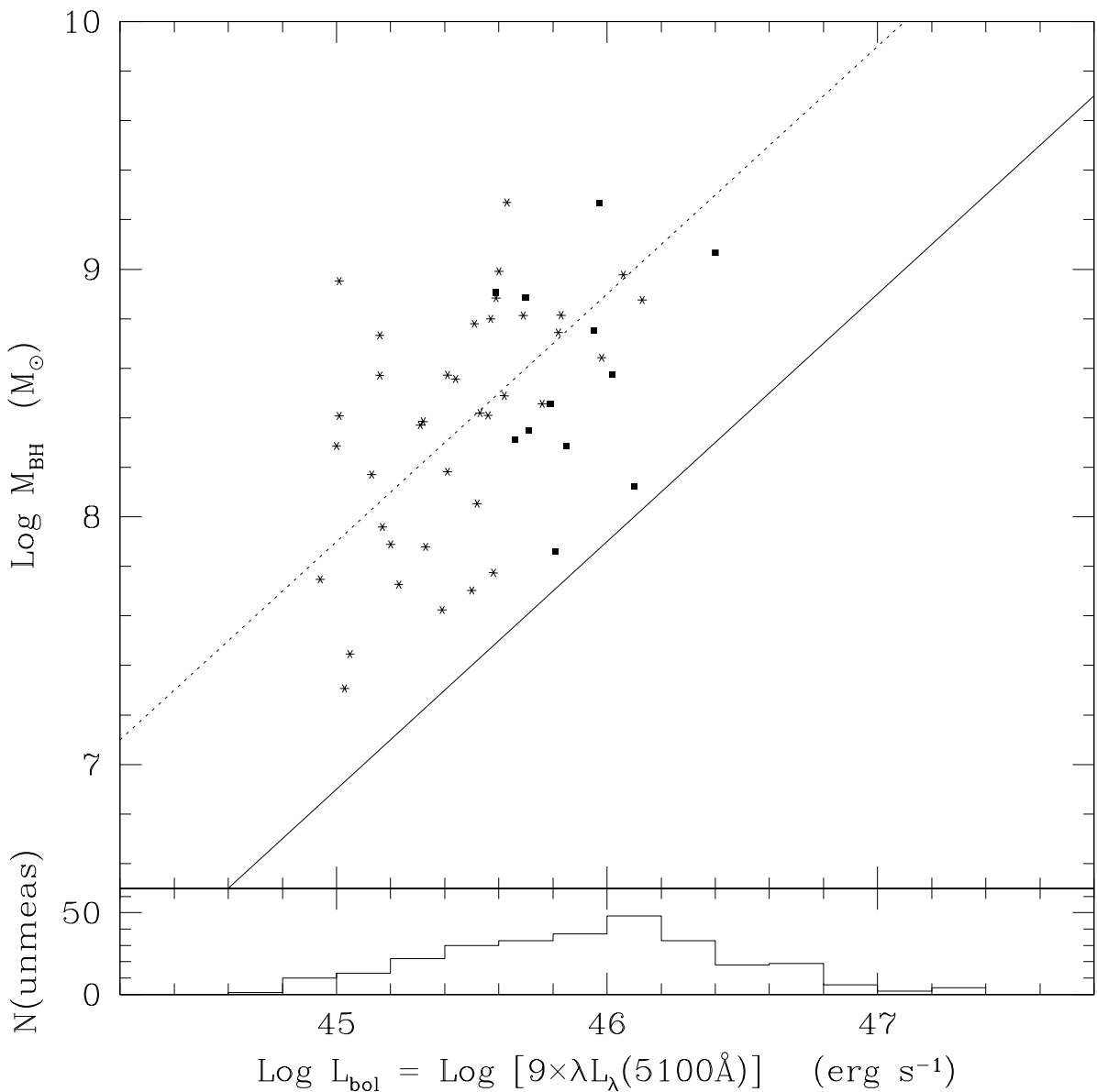} 
}
\caption{Estimated BH masses as a function of AGN bolometric luminosity.
  Points show objects eliminated from our final sample in cases for which
  the S/N of our data was insufficient to make a reliable FWHM measurement.
  Point types are as in Figure~\ref{fig:seurat}.  The histogram below shows
  the luminosity distribution of objects for which no sensible FWHM
  determination could be made (independent of the quality of the spectrum).
}
\label{fig:seuratbw}
\end{figure*}

\begin{figure*}
\centerline{
\epsfxsize=6.5truein
\epsfbox[41 219 508 688]{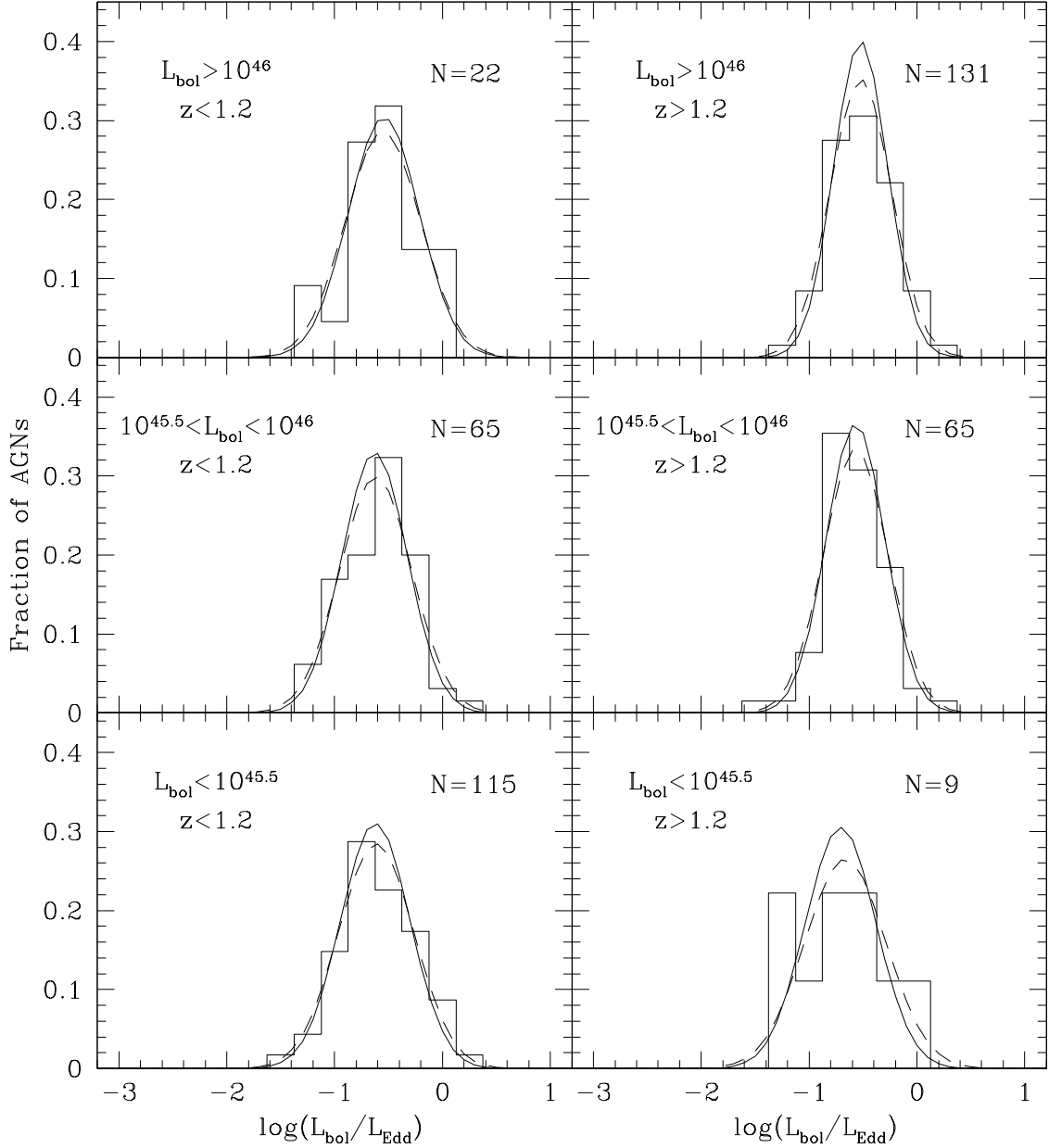}
}
\caption{Distributions of Eddington ratios in bins of luminosity and
  redshift. The panels are divided between $z<1.2$ (left) and $z>1.2$
  (right) and in increasing luminosity from bottom to top.  The histograms
  show the fraction of data points at each value of $L_{bol}/L_\edd$.
  Dashed curves are Gaussians with the same mean and dispersion as the data.
  Solid curves are the best-fit Gaussians, accounting for measurement
  uncertainties in luminosity and linewidth.  These curves are similar
  because the measurement errors are small compared to the distribution
  widths.  At all redshifts and luminosities, we find that most AGNs are
  radiating close to Eddington, with a dispersion of only $\sim 0.3$~dex.  }
\label{fig:lumbin}
\end{figure*}

\begin{figure*}
\centerline{
\epsfxsize=6.5truein
\epsfbox[18 146 581 703]{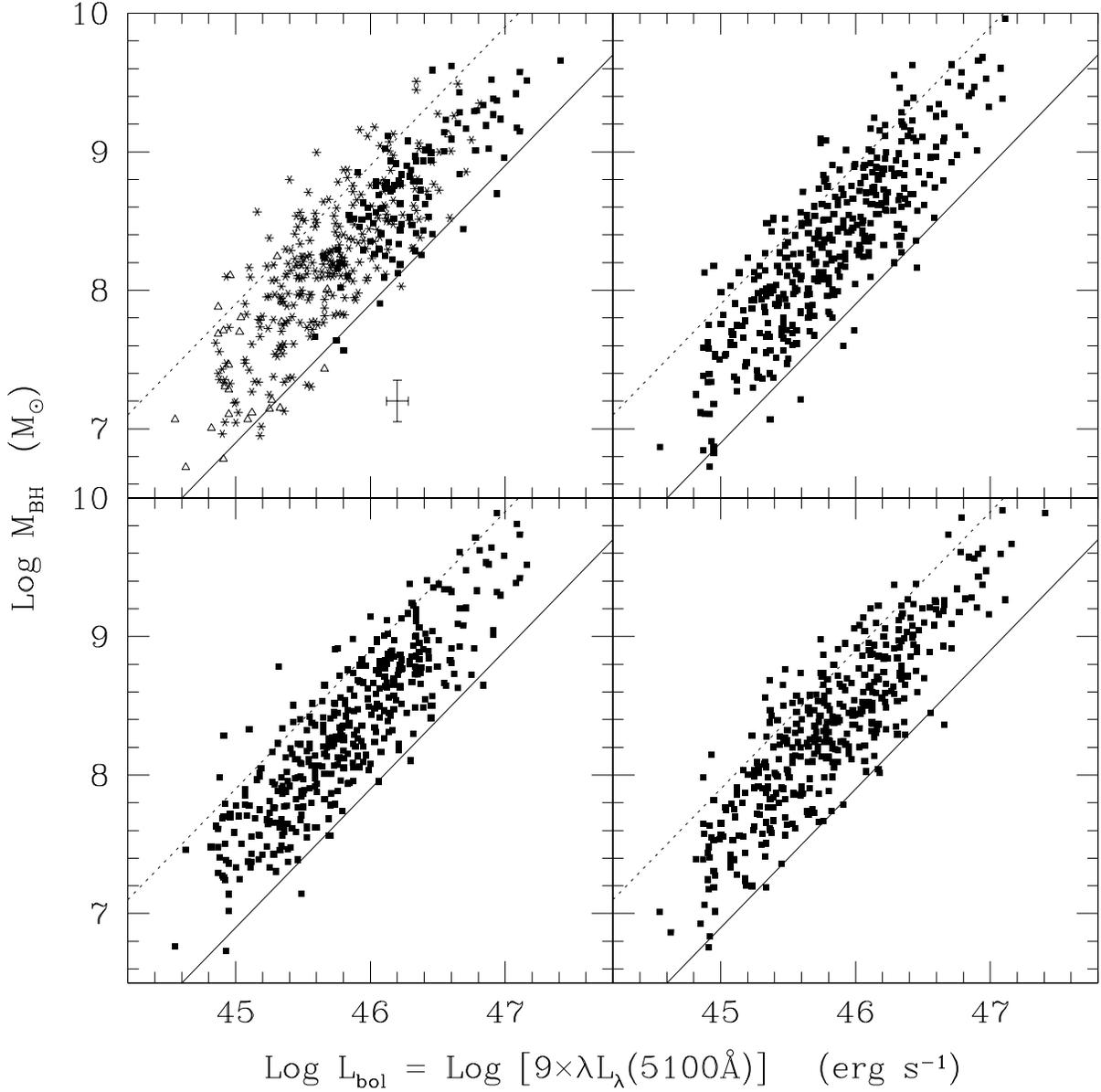}
}
\caption{Monte Carlo realizations of BH mass vs. bolometric luminosity.  The
  upper left panel shows the actual data, repeated from
  Fig.~\ref{fig:seurat}.  The remaining panels are 3 different Monte Carlo
  realizations of our data in which we retain our observed luminosities and
  draw the $\log(L_\bol/L_\edd)$ (and hence $M_{\bh}$) value for each AGN
  from a single Gaussian distribution with mean $\mu=-0.6$ and dispersion
  $\sigma=0.3$.  The distribution of points in the simulated datasets
  supports our inference from Fig.~\ref{fig:lumbin}, that a single, narrow
  Gaussian is an adequate description of the data.}
\label{fig:monte_lumbin}
\end{figure*}

\begin{figure*}
\centerline{ 
\epsfxsize=6.0truein 
\epsfbox[18 146 581 703]{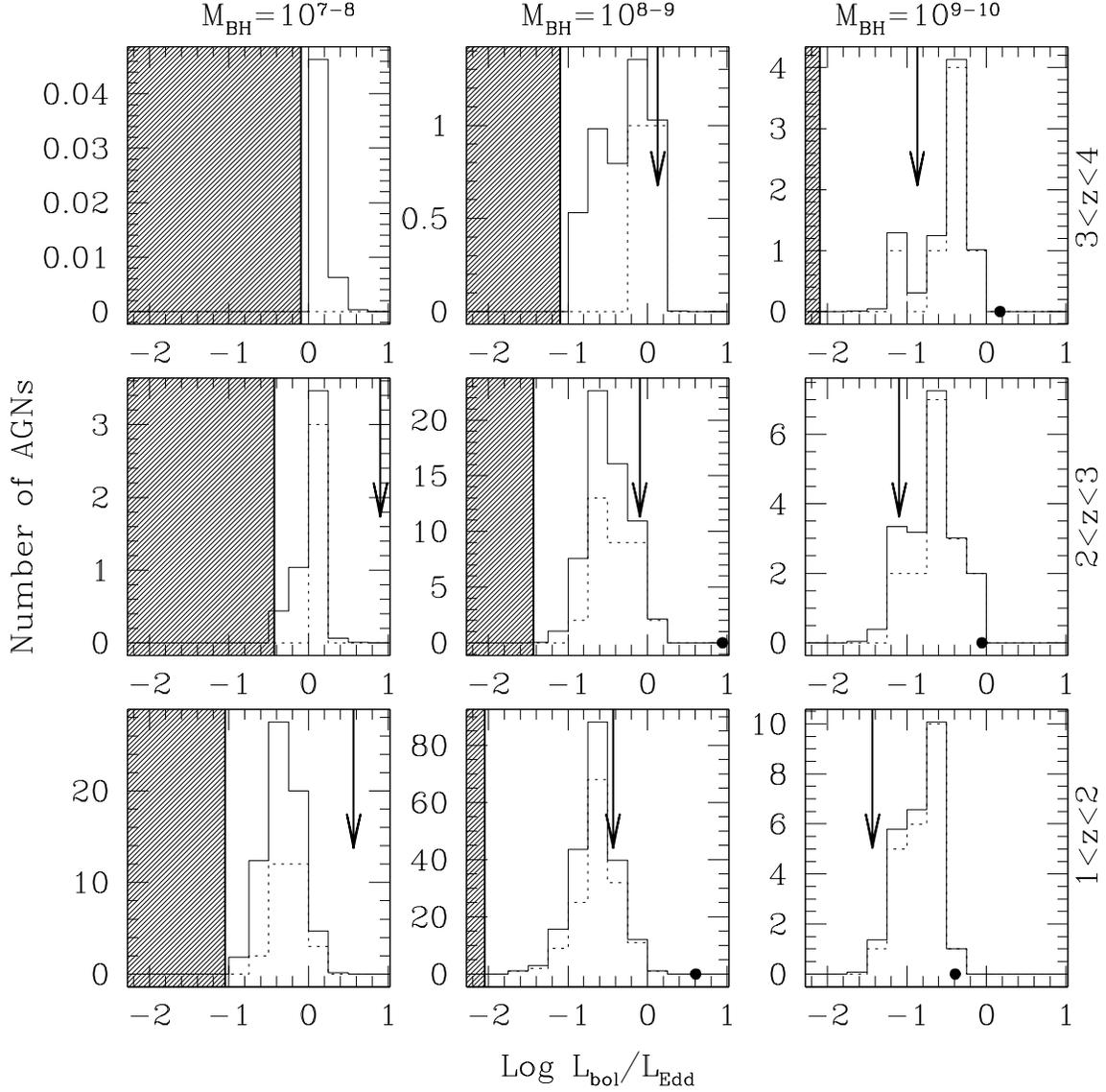} 
}
\caption{Distribution of Eddington ratios in bins of fixed BH mass and
  redshift.  Panels from left to right show 3 bins in mass, each 1 decade
  wide, ranging over $M_{\rm BH}=10^{7-10} M_{\odot}$. Panels from top to
  bottom show the redshift ranges $3<z<4$, $2<z<3$, and $1<z<2$.  The solid
  histograms in each panel show the distributions for the ``clean'' dataset
  of 407 objects corrected for completeness.  The dotted histograms show our
  data prior to the completeness-correction.  The arrows show where AGNs
  within the bin are first hitting the optical flux limit.  The shaded
  regions mark where AGNs are completely lost to optical selection. The
  position of the solid dots along the bottom of each panel show the
  equivalent of the arrows for the SDSS spectroscopic flux limit for that bin (in
  the lowest mass bins, these lie to the right of the plotted $x$-axis
  range).  }
\label{fig:massbin}
\end{figure*}

\begin{figure*}
\centerline{ 
\epsfxsize=6.5truein 
\epsfbox[18 146 581 703]{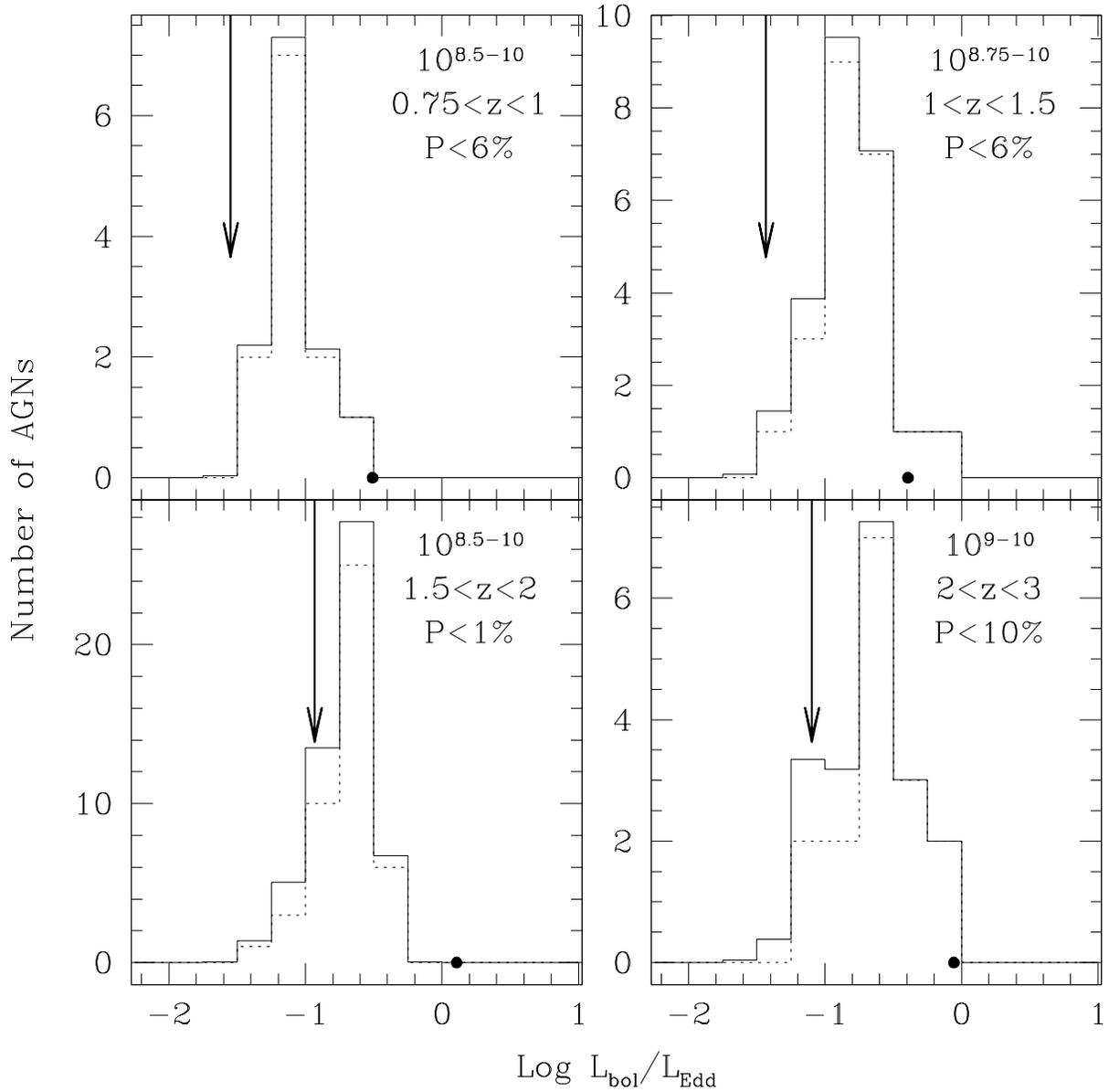} 
}
\caption{Eddington ratio distribution at fixed mass in bins for which
  distribution shape is not determined by optical selection.  The mass and
  redshift range for the bin is designated within each panel.  Solid
  histograms show our completeness-corrected values and dotted lines
  correspond to our raw measurements.  The arrows show where AGNs within the
  bin are first hitting the optical flux limit.  The solid dots along the
  bottom of each panel show the equivalent of the arrows for the SDSS
  spectroscopic limit for that bin.  The Poisson probabilities listed in
  each panel reflects the chance that the peak height of the solid
  histograms is produced by small number statistics alone.}
\label{fig:massbingood}
\end{figure*}

\begin{figure*}
\centerline{ 
\epsfxsize=6.5truein 
\epsfbox[18 146 581 703]{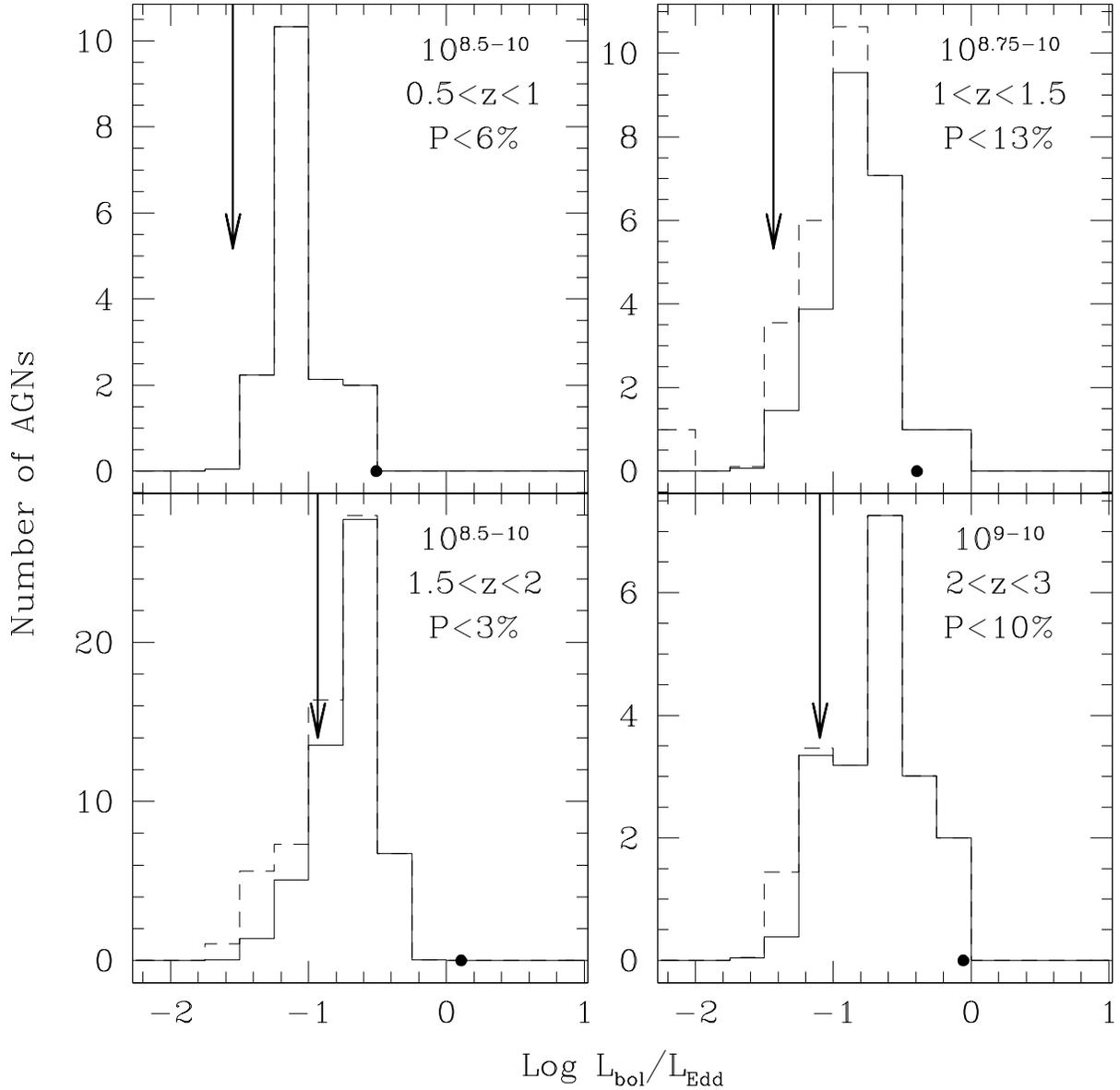} 
}
\caption{Effect of object removal on the peak in Eddington ratio
  distribution at fixed mass in bins for which distribution shape is not
  determined by optical selection.  The mass and redshift range for each bin
  is designated within each panel.  Solid lines show our
  completeness-corrected measurement as in Fig.~\ref{fig:massbingood} and
  dashed histograms include objects removed for low S/N (also corrected for
  completeness).  The arrows show where AGNs within the bin are first
  hitting the survey optical flux limit.  The solid dots along the bottom of
  each panel shows the equivalent of the arrows for the SDSS spectroscopic flux
  limit for that bin. The Poisson probabilities listed in each panel
  reflects the chance that the peak height of the dashed histograms is
  produced by small number statistics alone.}
\label{fig:massbingood2}
\end{figure*}

Our analysis does not test the zero-point or luminosity scaling of the
$M_\bh$ relations.  To aid comparisons with previous work, we have
used the scaling relations of \cite{mcj02} for H$\beta$ and
\cite{vest04} for \ion{C}{4}.  The results of \cite{onken04}, in which
reverberation-based BH masses were calibrated by assuming that AGNs
follow the same correlation between BH mass and stellar velocity
dispersion as quiescent galaxies, suggest that the zero points of
these relations should be adjusted upward by factors of 1.4 and 1.8,
respectively.  We did adjust the \cite{mcj02} \ion{Mg}{2} relation
based on the empirical evidence for a different luminosity scaling, as
shown in Figure~\ref{fig:dm.l}.  Altering the zero-points of the
$M_\bh$ relations or the mean bolometric correction would change our
conclusions about the {\it location} of peaks in the $L_\bol/L_\edd$
distributions, but they would not change our conclusions about the
{\it existence} of these peaks.  For example, the \cite{onken04}
revision would shift the means of our $L_\bol/L_\edd$ distributions
downward by $\sim 0.2$~dex, but would not broaden the distributions
(this shift has been incorporated into the analysis of
\citealt{vest06}.) With an identical shift to the mass ranges in
Figure~\ref{fig:massbingood}, the relative positions of the histograms
and the arrows in each panel would remain the same, but everything
would be at a slightly lower Eddington ratio.  Given the numerous
complexities of BH mass estimates from emission line widths, our
empirical evidence that these estimates have less than 0.3~dex rms
scatter is remarkable.  One might expect scatter nearly this large
from geometrical effects alone, since the relation between observed
linewidth and the BH mass may depend on viewing angle.

The dominance of the $z>0.5$ AGN population by near-Eddington accretors is
very different from the local-universe sample of AGNs, for which the
distribution of inferred Eddington ratios is broader
(e.g.,\citealt{woo02,ho04,heckman04}).  The luminosities of these local AGNs
are typically much lower than those in our sample, so it is unclear whether
luminosity or redshift dependence is responsible for most of the difference.
The AGES-I sample does not probe down to Seyfert luminosities (below $\sim
10^{44}\; {\rm erg}\; {\rm s}^{-1}$), and we therefore cannot comment on the
contribution of broad-lined Seyfert galaxies to the overall black hole
growth.  In addition, while narrow-line objects cannot be analyzed using the
methods employed here, they are a very small fraction of the overall AGES-I
AGN sample, with only 29 objects in the range $0.5\leq z\leq 1$ matching a
narrow-line template and no such objects above $z=1$.  High Eddington ratios
for the most luminous quasars at high redshift are not surprising, since the
steep high-mass fall-off of the BH mass function (see, e.g.,
\citealt{aller02}) means that the BHs required to power them at {\it
  sub}-Eddington luminosities would be very rare.  However, the dominance of
near-Eddington accretors at the knee of the luminosity function and at
redshifts at which the quasar luminosity function is declining in amplitude
imposes strong constraints on the distribution of BH fueling rates
\citep{steed03}.  Of course, there must also be many BHs with much lower
$L_\bol/L_\edd$ (perhaps approaching zero), otherwise it would be hard to
reconcile the local BH mass function with the observed number of AGNs.
However, these low Eddington-ratio objects do not contribute significantly
to the observed AGN population in the luminosity and redshift range probed
by AGES-I.

For high mass BHs at $z>1$, we can make these constraints more direct
by computing the $L_\bol/L_\edd$ distributions at fixed BH mass.  The
depth of the AGES-I survey is crucial in allowing us to compute
histograms for $L_\bol/L_\edd > 0.1$ that are unaffected by the survey
magnitude limit (see Fig.~\ref{fig:massbingood}).  Accounting for
completeness corrections, these histograms clearly decline toward both
high and low $L_\bol/L_\edd$ from the peak.  This peak could be
further studied by the expansion of the AGES survey presently underway
(AGES-II) and the 2SLAQ survey, which probes to approximately the same
depth as AGES-I over a much wider area \citep{richards05}.  However,
because 2SLAQ AGNs are selected by optical colors, the effects of dust
extinction may be more severe for this survey than for AGES.

Our results strongly suggest that BHs gain most of their mass while
accreting at near-Eddington rates.  Since the height of the Eddington ratio
distribution falls as $L_\bol/L_\edd$ drops from 0.3 to 0.1, any rise at
lower values would have to be extremely steep to contribute more mass growth
than the observed peak.  The main loopholes are the possibilities that much
of the growth occurs in objects that are obscured at optical wavelengths or
are accreting with very low efficiency.  Such objects could plausibly have a
different $L_\bol/L_\edd$ distribution.  However, the reasonable agreement
between the integrated emissivity of the optical quasar population and the
local BH mass density (\citealt{soltan82}; see \citealt{shankar04} for a
recent analysis) suggests that any such population cannot dominate by a
large fraction.  In fact, the \citet{shankar04} analysis concluded that
Eddington ratios of $\sim 1/3$ at $z=3$ were required in order to obtain a
good match between the emissivity from optical quasars and the local BH mass
density.  We also note that our AGN sample is not as adversely affected by
obscuration effects as most existing surveys.  While the X-ray-selected AGNs
are affected by soft X-ray absorption, the 24$\micron$-selected AGNs are
not.  Furthermore, the bluest optical band used for target selection in this
survey is the $R$-band instead of the more commonly used optical $B$-band.
This relative immunity to obscuration will be enhanced further in the
AGES-II AGN sample, which will have an optical flux limit ($I = 21.5$ mag)
that is both at a longer wavelength and deeper.

For a BH to become active, some event (merger, tidal interaction, dynamical
instability, etc.) must drive material into the central few parsecs of its
host galaxy, and this inner material must then form an accretion flow onto
the BH itself, on the much smaller scale of hundreds to thousands of AU.
The galactic-scale fueling events are likely to have a broad mass
distribution, and there is no reason for them to know about the precise mass
of the central BH.  The sharp peak of the observed $L_\bol/L_\edd$
distribution suggests that these events are often sufficient to provide
super-Eddington fuel supplies and that the actual BH accretion rates are
determined by the BH's self-regulation of the inner accretion flow.  The
narrow width ($\leq 0.3$~dex) and central value ($L_\bol/L_\edd\sim 1/4$)
are important targets for theoretical models of accretion flows, though
further investigation of the zero-point calibration of $M_\bh$ indicators is
desirable to firm up the latter constraint.  Overall, the population of
active BHs in the AGES-I survey is simpler than one might have imagined
beforehand, and explaining this simplicity is a new challenge for theories
of AGN evolution.

\acknowledgments

We thank Brad Peterson for providing access to his line-measuring routine.
We thank him and Rick Pogge, Marianne Vestergaard, Smita Mathur, Pat Osmer,
Casey Watson, and Kate Brand for helpful suggestions, detailed comments, and
advice.  We also thank Michael Brown, Nelson Caldwell, Dan Fabricant, Paul
Green, and Christine Jones for making the AGES project a success.  Finally,
we thank the referee for a helpful and thorough report.

Work by JAK and AG was supported by grant AST-0452758 from the NSF.  CAO
acknowledges The Ohio State University for support through the Distinguished
University Fellowship. Any opinions, findings, and conclusions or
recommendations expressed in this material are those of the authors and do
not necessarily reflect the views of the National Science Foundation.  This
work made use of images and/or data products provided by the NOAO Deep
Wide-Field Survey (Jannuzi and Dey 1999), which is supported by the National
Optical Astronomy Observatory (NOAO).  NOAO is operated by AURA, Inc., under
a cooperative agreement with the National Science Foundation.

\clearpage

\begin{deluxetable}{ccccccccc}
\tablewidth{0pt} \tablecaption{Gaussian Parameters of Data and Fits to
Data} \tablehead{\colhead{$z_{\rm bin}$\tablenotemark{a}} &
\colhead{$L_{\rm bin}$\tablenotemark{b}} & \colhead{$N$} &
\colhead{$\mu_{\rm data}$\tablenotemark{c}} & \colhead{$\sigma_{\rm
data}$ } & \colhead{$S_k$ \tablenotemark{d}} & \colhead{$A_k$
\tablenotemark{e}} & \colhead{$\mu_{\rm model}$} &
\colhead{$\sigma_{\rm model}$}} 
\startdata 
Low & Low & 115 & $-$0.61 & 0.35 & $\ \ $0.01 & 2.69 & $-$0.62 & 0.32 \\ 
High & Low & 9 & $-$0.66 & 0.38 & $\ \ $0.08 & 1.78 & $-$0.70 & 0.32 \\ 
Low & Med & 65 & $-$0.62 & 0.33 & $-$0.07 & 2.49 & $-$0.63 & 0.30 \\ 
High & Med & 65 & $-$0.56 & 0.29 & $-$0.17 & 3.77 & $-$0.57 & 0.27 \\ 
Low & High & 22 & $-$0.55 & 0.34 & $-$0.27 & 2.55 & $-$0.54 & 0.33 \\ 
High & High & 131 & $-$0.52 & 0.28 & $-$0.02 & 2.87 & $-$0.52 & 0.24 \\ 
\enddata
\tablenotetext{a}{ Low: $z<1.2$; High: $z>1.2$} \tablenotetext{b}{ Low:
$\log L_\bol < 45.5$; Med: $45.5 < \log L_\bol < 46$;\\ High:~$\log
L_\bol > 46$}
\tablenotetext{c}{ $\sigma_{\mu}=\sigma_{data}/\sqrt{(N)}$}
\tablenotetext{d}{ $\sigma_{S_k}=\sqrt{(6/N)}$}
\tablenotetext{e}{ $\sigma_{A_k}=\sqrt{(24/N)}$} 
\tablecomments{For each bin in redshift, $z_{\rm bin}$, and luminosity, 
$L_{\rm bin}$, the table lists the number of objects, $N$, the mean 
Eddington ratio, $\mu_{\rm data}$, the dispersion in Eddington ratios, 
$\sigma_{\rm data}$, the skewness of the distribution, $S_{k}$, the 
kurtosis of the distribution, $A_{k}$, the maximum likelihood fit to the 
mean Eddington ratio, $\mu_{\rm model}$, and the maximum likelihood fit to 
the dispersion in Eddington ratios, $\sigma_{\rm model}$. In addition, we
give the formulae for the errors in the mean, skew, and kurtosis.}
\label{tab:mgauss} 
\end{deluxetable}

\clearpage

\end{document}